\documentclass{vldb}
\usepackage{xcolor}
\usepackage{graphicx}
\usepackage{float}
\usepackage{subfig}
\usepackage{xspace}
\usepackage{textcomp}
\usepackage{times}
\usepackage{mathptmx}
\usepackage[font={it}]{caption}
\usepackage[T1]{fontenc}
\usepackage{listings}
\usepackage{balance}
\usepackage{algorithm}
\usepackage[noend]{algpseudocode}
\usepackage{hyperref}
\usepackage{amsmath}
\usepackage{paralist}
\usepackage{soul}
\usepackage[normalem]{ulem}

\newcommand{\stitle}[1]{\smallskip \noindent \textbf{#1.}}
\newcommand{\emtitle}[1]{\smallskip \noindent \underline{\emph{#1.}}}

\newcommand{\prev}[1]{\noindent{}}

\newcommand{\papertext}[1]{}
\newcommand{\techreport}[1]{#1}

\newcommand{\later}[1]{}

\newcommand{\ntail}{{\sc NeedleTail}\xspace}

\newcommand{\densitymap}{{\sc DensityMap}\xspace}
\newcommand{\densitymaps}{{\sc DensityMap}s\xspace}
\newcommand{\lossy}{{\sc Lossy-Bitmap}\xspace}

\newcommand{\anyk}{any-$k$\xspace}
\newcommand{\Anyk}{Any-$k$\xspace}
\newcommand{\thresh}{{\sc Density-Optimal}\xspace}

\newcommand{\optimal}{{\sc IO-Optimal}\xspace}
\newcommand{\prong}{{\sc Locality-Optimal}\xspace}

\newcommand{\scan}{{\sc Disk-Scan}\xspace}
\newcommand{\bmscan}{{\sc Bitmap-Scan}\xspace}
\newcommand{\bmrand}{{\sc Bitmap-Random}\xspace}
\newcommand{\ewah}{{\sc EWAH}\xspace}
\newcommand{\hybrid}{{\sc Two-Phase}\xspace}
\newcommand{\hybridactual}{{\sc Hybrid}\xspace}

\newcommand{\shared}{\textsc{Shared-Scan}\xspace}
\newcommand{\bitcomb}{\textsc{Bitmap-Combined}\xspace}
\newcommand{\denscomb}{\textsc{Density-Combined}\xspace}

\DeclareMathOperator*{\argmax}{argmax}

\newtheorem{theorem}{Theorem}
\newtheorem{example}{Example}

\newtheorem{problem}{Problem}
\newtheorem{observation}{Observation}

\makeatletter
\def\@copyrightspace{\relax}
\makeatother

\newenvironment{denselist}{
    \begin{list}{\small{$\bullet$}}%
    {\setlength{\itemsep}{0ex} \setlength{\topsep}{0ex}
    \setlength{\parsep}{0pt} \setlength{\itemindent}{0pt}
    \setlength{\leftmargin}{1.5em}
    \setlength{\partopsep}{0pt}}}%
    {\end{list}}

\title{Optimally Leveraging Density and Locality \\ to Support LIMIT Queries}

\makeatletter
\algrenewcommand\ALG@beginalgorithmic{\small}
\makeatother

\begin{document}

\numberofauthors{1}
\author{
\alignauthor
\large \hspace{-15pt}Albert Kim$^{1\thanks{\small Both authors contributed equally to this work.}}$, Liqi Xu$^{2\footnotemark[1]}$, Tarique Siddiqui$^2$, Silu Huang$^2$, Samuel Madden$^1$, Aditya Parameswaran$^2$ \\
\affaddr{$^1$MIT \hspace{11em} $^2$University of Illinois (UIUC)} \\
\affaddr{\{alkim,madden\}@csail.mit.edu} \;
\affaddr{\{liqixu2,tsiddiq2,shuang86,adityagp\}}@illinois.edu
}

\maketitle

\begin{abstract}

Existing database systems are not optimized 
for queries with a LIMIT clause---operating
instead in an all-or-nothing manner.
In this paper, we propose a fast LIMIT query evaluation engine, called
\ntail, aimed at letting analysts browse a small
sample of the query results on large datasets
as quickly as possible, independent of the overall
size of the result set.
\ntail introduces {\em density maps}, a lightweight
in-memory indexing structure, and a set of 
efficient algorithms (with desirable theoretical guarantees)
to quickly locate promising blocks, 
trading off locality and density. 
In settings where the samples are used to compute aggregates,
we extend techniques from survey sampling
to mitigate the bias in our samples.
Our experimental results demonstrate that 
\ntail returns results 
 4$\times$ faster on HDDs and 9$\times$ faster on SSDs on average,
while occupying up to 23$\times$ less memory 
than existing techniques.

\end{abstract}


\section{Introduction}


Users of databases are frequently interested in retrieving only a small subset of records that satisfy a query, e.g., by specifying a LIMIT clause in a SQL query.  Although there has been extensive work on {\it top-k} queries that retrieve the largest or smallest results in a result set, or that tries to provide a random sample of results, there has been relatively little work on so-called {\it \anyk} queries that simply retrieve a few results without any requirements about the ordering or randomness of the results.

\Anyk has many applications in exploratory data analysis.
When exploring new or unfamiliar datasets,
users often issue arbitrary queries, 
examine a subset or sample (a ``screenful'') of records,
and incrementally and re-issue their queries
based on observations from this sample~\cite{tukey1977exploratory,hanrahan2012analytic}. 
To support this form of exploratory {\em browsing},
many popular SQL IDEs implicitly 
limit the number of result records displayed, recognizing
the fact that users often do not need to,
nor do they have the time to see all of the result records.
For example, 
phpMyAdmin\footnote{\small \url{docs.phpmyadmin.net/en/latest/config.html}} 
for MySQL has a \textsf{MaxRows} configuration
parameter;
pgAdmin\footnote{\small \url{pgadmin.org/docs/pgadmin3/1.22/query.html}}
for PostgreSQL has a \textsf{rowset size} parameter; and
SQL Developer\footnote{\small \url{oracle.com/technetwork/developer-tools/sql-developer/}}
for Oracle has a \textsf{array fetch size} parameter. 
Even outside of the context of SQL IDEs,
in tabular interfaces such as 
Microsoft Excel or Tableau's Table View,
users only browse or examine a ``screenful'' of records at a time.

LIMIT clauses provide a way to express \anyk in traditional data-bases, and are
 supported by most modern database systems~\cite{limit-stackoverflow}.
Surprisingly, despite their fundamental importance,
there has been little work in executing
LIMIT queries efficiently.
Existing databases support LIMIT/\anyk
by simply executing the entire query, and 
returning the first $k$ result records 
as soon as they are ready, via pipelining. 
This is often not interactive on large datasets,
especially if the query involves selective predicates.
To address this limitation, in this paper, we develop
methods that allow us to quickly identify a 
subset of records for arbitrary
\anyk queries.

Specifically, we develop indexing and query evaluation tech\-niq\-ues
to address the \anyk problem.
Our new data exploration engine, \ntail\footnote{\small We named \ntail 
after the world's fastest
bird~\cite{needletail}.},
employs 
{\em (i)} a lightweight
indexing structure, {\em density maps}, tailored for \anyk,
along with {\em (ii)} efficient algorithms that operate on 
density maps and select a sequence of data blocks that are 
{\em optimal for locality} (i.e., how close are the blocks to each other),
or {\em optimal for density} (i.e., how dense are the blocks in terms of
containing relevant results).
We couple these algorithms with {\em (iii)}
an algorithm that is {\em optimal for overall I/O}, by 
employing a {\em simple disk model},
as well as a {\em hybrid algorithm} that selects between the locality
and density-optimal variants. 
Finally, while \anyk is targeted at browsing,
to allow the retrieved results to be used for a broader 
range of use-cases involving aggregation (e.g., for computing statistics or visualizations), 
{\em (iv)} we extend {\em statistical survey sampling} techniques
to eliminate the bias in the retrieved results. 

We now describe these contributions and the underlying challenges 
in more detail.

\emtitle{(i) Density Maps: A Lightweight Indexing Scheme}
Inspired by bit\-map indexes, which are effective for
arbitrary read-only queries, but are somewhat expensive to store,
we develop a lightweight 
indexing scheme called \emph{density maps}.
Like bitmaps, we store a density map for each value of each attribute.
However, unlike bitmaps, which store a column of bits for each distinct value of an attribute,
density maps store an array of densities for each {\it block} of records, where the array contains
one entry per distinct value of the attribute. 
This representation allows density maps to be 3-4 orders of magnitude smaller than bitmaps, so we can easily store a density map for
every attribute in the data set, and these 
can comfortably fit in memory even on large datasets. 
Note that prior work has identified  other 
lightweight statistics that help rule out 
blocks that are not relevant for certain queries~\cite{wiki:lossybitmap,moerkotte2008small,lang2016data, boncz2005monetdb, francisco2011netezza}:
we compare against the appropriate ones in our experiments
and argue why they are unsuitable for \anyk
in Section~\ref{sec:related}.

\emtitle{(ii)-(iii) Density Maps: Density vs.~Locality}  In order to use density maps to
retrieve $k$  records that satisfy query constraints, one approach would be to
identify blocks that are likely to be ``dense'' in that they  contain more
records that satisfy the conditions and preferentially retrieve those blocks.
However, this {\em density-based} approach may lead to excessive random
access. Because random accesses---whether in memory, on flash, or on
rotating disks---are generally orders of magnitude slower than sequential
accesses, this is not desirable. Another approach would be to identify a
sequence of consecutive blocks with $k$ records that satisfy the conditions,
and take advantage of the fact that sequential  access is faster than random
access,  exploiting {\em locality}.  In this paper, we develop algorithms that 
are optimal from the perspective of density (called \thresh) and 
from the perspective of locality (called \prong).

\smallskip
\noindent
Overall, while we would like to optimize
for both density and locality, optimizing for one usually comes at
the cost of the other, so developing the globally optimal strategy to retrieve
$k$ records is non-trivial.
To combine the benefits of density and locality, we need 
a cost model for the storage media that can help us reason about
their relative benefits. 
We develop a simple cost model in this paper, and use this to develop
an algorithm that is optimal
from the perspective of overall I/O (called \optimal).  
We further extend the density and locality-optimal algorithms
to develop a hybrid algorithm (called \hybrid) 
that fuses the benefits of both approaches. 
We integrated all four of these algorithms, coupled with
indexing structures, into our \ntail data exploration engine. On both
synthetic and real datasets, we observe that {\em \ntail can be  several orders of
magnitude faster than existing approaches} when returning $k$ records that
satisfy user conditions. 

\emtitle{(iv) Aggregate Estimation with \Anyk Results}
In some cases, 
instead of just optimizing for retrieving \anyk,
it may be important to use the retrieved results to 
estimate some aggregate value. 
Although \ntail
can retrieve more samples in the same time,
the estimate of the aggregate may be biased,
since \ntail may preferentially
sample from certain blocks.
This is especially true 
if the attribute being aggregated
is correlated with the layout of data on disk or in memory. 
We employ {\em survey 
sampling}~\cite{horvitz1952generalization,lohr2009sampling}
techniques to support accurate aggregate estimation while retrieving \anyk records.
We adapt {\em cluster sampling} techniques to reason about
block-level sampling, along with {\em unequal probability estimation}
techniques
to correct for the bias.
With these changes, \ntail is able to achieve error rates similar to 
pure random sampling---our gold standard---in 
{\em much less time}, while 
{\em returning multiple orders of magnitude more records}
for the analyst to browse.
Thus, even when computing aggregates, \ntail is
substantially better than other schemes.

\smallskip
\noindent
Of course, there are other  techniques for improving analytical query response time,
including in-memory caching, materialized views,
precomputation, and materialized samples.  These techniques can also be applied to \anyk,
and are largely orthogonal to our work.  One advantage of our density-map approach versus much of this related work is
that it does not assume a query workload is available or that queries are predictable, 
which enables us to support truly exploratory data analysis.
Similarly, traditional indexing
structures, such as B+ Trees, could efficiently answer some \anyk queries.
However, to support \anyk queries with arbitrary predicates,
we would need B+ Trees on every
single attribute or combination of attributes, 
which often will be prohibitive in terms of space. Bitmap indexes are a more space
efficient approach, but even so, 
storing a bitmap in memory
for every single value for every single attribute (10s of values
for 100s of attributes) is impossible for large datasets, as we show in our
experimental analysis.

\stitle{Contributions and Outline}
The chief contribution of this paper is the design and development
of \ntail, an efficient 
data exploration engine for both browsing and aggregate estimation 
that retrieves samples in orders of magnitude faster than other approaches.
\ntail's design includes its density map indexing structure,
retrieval algorithms (\thresh, \prong, \hybrid, and \optimal) with extensions for complex queries,
and statistical techniques to correct for biased
sampling in aggregate estimation.

We formalize the browsing problem in Section~\ref{sec:prob},
describe the indexing structures in Section~\ref{sec:idx},
the \anyk sampling algorithms in Section~\ref{sec:alg} and~\ref{sec:algo_hybrid},
the statistical debiasing techniques in Section~\ref{sec:est},
extensions to complex queries in Section~\ref{sec:join},
and the system architecture in Section~\ref{sec:arch}.
We evaluate \ntail in Section~\ref{sec:eval}.

\section{Problem Formulation}
\label{sec:prob}

We now formally define the \anyk problem.
We consider a
standard OLAP data exploration setting where we have a database
$D$ with a star schema consisting of continuous measure attributes $M$ and
categorical dimension attributes $A$. 
For simplicity, 
we focus on a single database table $T$, with $r$
dimension attributes and $s$ measure attributes, leading to the schema: $T
= \{A_1, A_2, ..., A_r, M_1, M_2, ..., M_s\}$; our techniques generalize beyond 
this case, as we will show in later sections. 
We
use $\delta_i$ to denote the number of distinct values for the dimension
attribute $A_i$ with distinct values $\{V_i^1, V_i^2,..,V_i^{\delta_i}\}$.

Consider a selection query $Q$ on $T$ where the selection condition is 
a boolean formula
formed out of equality predicates on
the dimension attributes $A$.
We define the set of records which form the result set of query $Q$ to be
the {\em valid records} with respect to $Q$.
As a concrete example,
consider a data analyst exploring campaign finance data. 
Suppose they want to find \anyk individuals who donated to Donald Trump, live in a
certain county, and are married. 
Here,  the query $Q$ on $T$ has a selection condition that is a conjunction
of three predicates---donated to Trump, lives in a particular county, and 
is married. 

Given the query $Q$, traditional databases would return
all the valid records for $Q$ in $T$, irrespective of how long it takes. 
Instead, we define an \anyk query
$Q_k$ as the query which returns $k$ valid records out of the set of all valid records 
for a given query $Q$. $Q_k$ can be
written as follows:
\vspace{-3pt}
\begin{align*}
  \text{SELECT } \text{\bf ANY-K}(*) \text{ FROM }T \text{ WHERE }\langle\text{CONDITION}\rangle
\vspace{-7pt}
\end{align*}
For now we consider simple selection
queries of the above form; we show how to extend 
our approach to support aggregates in Section~\ref{sec:est} 
and how to support grouping and joins in Section~\ref{sec:join}.

We  formally state the \anyk
sampling problem for simple selection queries as follows:
\vspace{-3pt}
\begin{problem}[\anyk sampling]
Given an \anyk query $Q_k$,
the goal of \anyk sampling is to
retrieve any $k$ valid records in as little time as possible.
\vspace{-3pt}
\end{problem}
Note that unlike random sampling, \anyk does not require
the returned records to be randomly selected. Instead, \anyk sampling prioritizes
query execution time over randomness.
We will revisit the issue of randomness in Section~\ref{sec:est}.
Next, we develop the
indexing structures required to support \anyk algorithms.
%
\later{
Alternatively, we can also define the problem with a constrained time limit:
\vspace{-3pt}
\begin{problem}[time-constrained \anyk sampling]
Given an \anyk query $Q_K$ on a table $T$ and a query execution time limit 
$\tau$, retrieve the maximum possible number of valid records.
\vspace{-3pt}
\end{problem}
We focus primarily on the first problem but describe extensions for the second 
problem when appropriate.
}


\section{Index Structure}
\label{sec:idx}
To support the fast retrieval of \anyk samples, we develop a lightweight indexing
structure called the \densitymap. \densitymaps share some 
similarities with bitmaps, so
we first briefly describe bitmap indexes.
We then discuss how \densitymaps address the
shortcomings of bitmap indexes.


\subsection{Bitmap Index: Background}
Bitmap indexes~\cite{o1989model} are commonly used for ad-hoc queries in
read-mostly
workloads~\cite{chan1998bitmap,Wu:SciDAC09:2009,DBLP:journals/tods/SinhaW07,DBLP:conf/ideas/WuMC10}. 
Typically, the index contains one bitmap
for each distinct value $V$ of each dimension attribute $A$ in a table.
Each bitmap is a vector of bits in which the $i$th bit is set to $1$, 
if $A=V$
for the
$i$th record, and 0 otherwise. 
If a query has a equality predicate on only one
attribute value, we can simply look at the corresponding bitmap for that
attribute value and return the records whose bits are set in the bitmap. For
queries that have more than one predicate, or range predicates, we must perform bitwise AND or OR
operations on those bitmaps before fetching the valid records. Bitwise
operations can be executed rapidly, particularly when bitmaps fit in memory.

Although bitmap indexes have proven to be effective for traditional OLAP-style
workloads,
these workloads typically consist of queries in which the user expects to
receive all valid records that match the filter. 
Nevertheless, bitmap indexes can be used for \anyk sampling.
One simple strategy would be to
perform the bitwise operations across all predicated bitmap indexes, perform a
scan on the resulting bitmap, and return the first $k$ records with matching bits.
However, the efficiency of this strategy greatly depends on the layout of the valid
records. For example, if all valid records are clustered near the end of the dataset,
the system would have to scan the entire bitmap index before finding the set
bits.  
\techreport{Furthermore, 
returning the first matching $k$ records may be sub-optimal
if the first $k$ records are dispersed across the dataset,
since retrieving each record would result 
in a random access.  
If a different set of $k$ records existed later in the dataset, 
but with better locality, a preferred strategy might be to return that second set
of records instead.}

In addition to some limitations when performing \anyk sampling, bitmap indexes often take
up a large amount of space, since we need to store one bitmap per distinct value of each dimension. 
As the number of attribute values and dimension
attributes increase, a greater number of bitmap indexes is required. 
Even with various methods to compress bitmaps, such as BBC~\cite{bbc},
WAH~\cite{wah}, PLWAH~\cite{plwah}, EWAH~\cite{lemire2010sorting}, density maps consume orders of magnitude less space
than bitmap indexes,  as we show in
Section~\ref{sec:eval}.

\subsection{Density Map Index}
We now describe how \densitymap
addresses the shortcomings of bitmap indexes.
Our design starts from the observation
 that modern hard disk drive (HDDs) typically have 4KB minimum storage units
called sectors, and systems may only read or write
from HDDs in whole sectors. Therefore, it takes the same amount of
time to retrieve a block of data as it does a single record.
\densitymaps take advantage of this fact to reason about the data at a 
block-level, rather than at the record-level as bitmaps do.  Similarly, SSD and RAM access pages or cache lines of data at a time.

Thus, for each block, 
a \densitymap stores the frequency of set bits in that
block, termed the \emph{density}, rather than enumerating the set bits.
This enables the system to 
``skip ahead''
to these dense blocks to retrieve the \anyk samples.  
Further, by storing block-level statistics rather than record-level statistics,
\densitymaps can greatly reduce the amount of indexing space required
compared to bitmaps. \techreport{In fact, \densitymaps can be thought of a form of
lossy compression.}
Overall, by storing density-based information at the block level, we
benefit from {\em smaller size} and {\em more relevant information} tailored
to the \anyk sampling problem.

Formally, for each attribute $A_i$, and each attribute value
taken on by $A_i$, $V_i^j$, we store a \densitymap $D_i^j$, consisting of
$\lambda$ entries, one corresponding to each block on disk. We can express
$D_i^j$ as $\{d_{i_1}^j,d_{i_2}^j,..,d_{i_k}^j,..,d_{i_\lambda}^j\},$ where
$d_{i_k}^j$ is the percentage of tuples in block $k$ that satisfy the
predicate $A_i=V_i^j$.
Note that while \densitymaps are stored per column, the actual underlying data is assumed
to be stored in row-oriented fashion.

\begin{figure}[h!]
\vspace{-10pt}
\includegraphics[width=\linewidth]{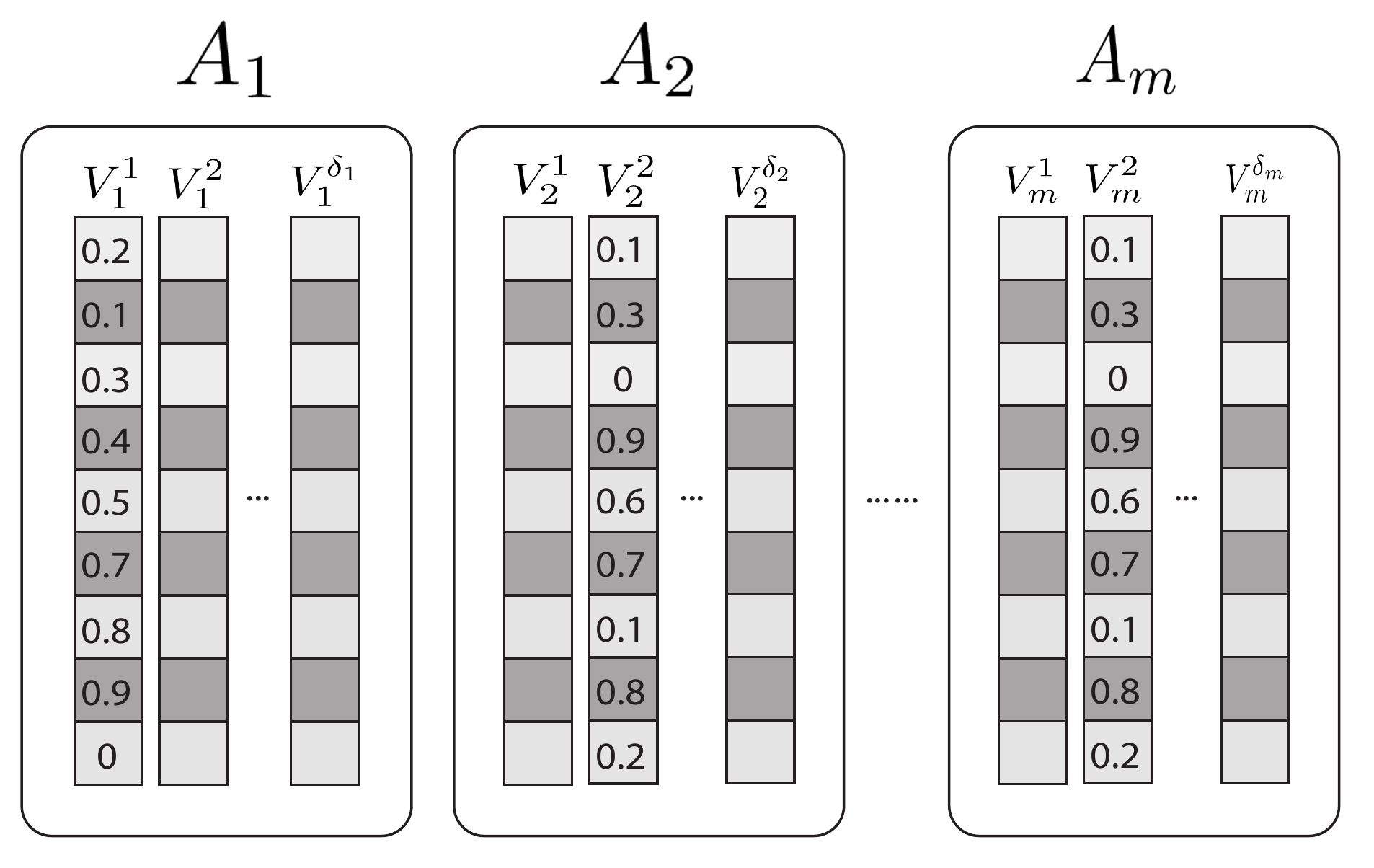}
\vspace{-20pt}
\caption{$\densitymap$s } 
\label{fig:densityMap}
\vspace{-12pt}
\end{figure}

\begin{example} 
The table in Figure~\ref{fig:densityMap} is stored over 9
blocks. The density map $D_1^1$ for $V_1^1$ is
$\{0.2,0.1,0.3,0.4,0.5,0.7,0.8,0.9,0\}$, indicating 20 percent of tuples in
block 1 and 10 percent of tuples in block 2 have value $V_1^1$ for attribute
$A_1$ respectively.
\end{example}

\densitymaps are a very flexible index structure as they can estimate the
percentage of valid records for any ad-hoc query with single or nested
selection constraints. For queries with more than one predicate, we can
combine multiple \densitymaps together to calculate the estimated percentage
of valid records per block, multiplying densities for conjunction and adding
them for disjunction. 
In performing 
this estimation, we implicitly assume that
the \densitymaps are independent, akin 
to selectivity estimation 
in query optimization~\cite{complete-book}. 
As in query optimization, this assumption 
may not always hold, but as we demonstrate in our experiments
on real datasets, it still leads to effective results.
In particular, \densitymaps drastically reduce
the number of disk accesses
by skipping blocks whose estimated densities are zero
(and thus definitely do not contain valid records).
Some readers may be reminded of other statistics
used for query optimization, such as histograms~\cite{complete-book}.
However, unlike histograms which store the overall frequencies of records for
entire relations,
%
\densitymaps
store this information at a finer block-level granularity.

\vspace{-5pt}
\begin{example} In Figure~\ref{fig:densityMap}, for a given query Q with
selection constraints $A_1$ = $V_1^1$ AND $A_2$ = $V_2^2$, the estimated
\densitymap after combing $D_1^1$ and $D_2^2$ is $\{0.02, 0.03, 0, 0.36, 0.3,
0.49, 0.08, 0.72, 0\}$, indicating (approximately) that block 1 has 2 percent matching records, and block 2 has 3 percent matching records for Q.
\vspace{-5pt}
\end{example}

Thus, compared to bitmaps, \densitymaps are a coarser statistical summary of
valid records in each block for each attribute value. 
\densitymaps save significant storage costs
by keeping information at the block-level instead of record-level, making
maintaining \densitymaps in memory feasible. Moreover, coupled with efficient
algorithms, which we describe in detail next,
\densitymaps can decrease the number of blocks read from disk for \anyk
sampling and therefore 
reduce the query execution time.

\techreport{One concern with \densitymap is that, since we admit all records from a block
which satisfy the constraints into our \anyk sample set, the samples we retrieve may
be biased with
respect to the data layout. \prev{However, later in the
paper, we empirically show that by retrieving high density blocks
first, we can fetch many more \anyk samples faster than random sampling;
thus, allowing us to reach similar error rates similar to true random sampling
with shorter response times using \densitymaps. 
Additionally, for aggregate queries, as described}
In Section 
\ref{sec:est}, we describe techniques to correct the bias due to possible correlations 
between the samples and the data layout by applying
cluster sampling and unequal probability estimation techniques.}




\section{Any-K Algorithms: Extremes}\label{sec:alg}

We introduce two algorithms which take advantage of our \densitymaps to
perform fast \anyk sampling. The primary
insights for these algorithms come from the following two observations.

First, a high density block has more valid records than a low density block. Thus, it
is more beneficial 
to retrieve the high density block,
so that overall, fewer blocks are retrieved.
\vspace{-3pt}
\begin{observation}[Density: Denser is better.]
Under the same circumstances, retrieving a block with high density is
preferable to retrieving a low density block.
\label{obs:density}
\end{observation}
\vspace{-3pt}

In a HDD, the time taken to retrieve a block from disk can be split into
\emph{seek time} and \emph{transfer time}. The seek time is the time it takes
to locate the desired block, and the transfer time is the time required to
actually transfer the bytes of data in the block from the HDD to
the operating system. Blocks which are far apart incur additional seek time,
while neighboring blocks typically only require transfer time. Thus, retrieving
neighboring blocks is preferred.  Similar locality arguments hold (to varying degrees) 
on SSD
and RAM.

\vspace{-3pt}
\begin{observation}[Locality: Closer is better]
Under the same circumstances, retrieving neighboring blocks is preferable to
retrieving blocks which are far apart.
\label{obs:locality}
\end{observation}
\vspace{-3pt}

Our basic \anyk sampling algorithms take advantage of each of these
observations: 
\thresh optimizes for density while \prong optimizes for locality.
These two algorithms are {\em optimal extremes}, favoring just one of locality or density.

On different types of storage media, the two observations can have different
amounts of impact. For example, the locality observation may not be as
important
for in-memory data and solid-state drives (SSDs), since the random I/O
performance of these storage media is not as poor as it is
on HDDs.
For our purposes, we focus on the HDDs, which is the most common type of storage device, 
but we also evaluate our techniques on SSDs.


To judge which of these two algorithms 
is better in a given setting, or to combine the benefits of these
two algorithms, we require a cost model for storage media, which we
present in Section~\ref{sec:algo_hybrid}.

Table~\ref{tab:notation} provides a summary of the notation used in the following sections.

\begin{table}[htbp]
\vspace{-10pt}
 \begin{center}
 \scalebox{0.75}{    
  \begin{tabular}{c|c} 
       Symbol & Meaning \\ \hline \hline 
       $\lambda$ & Number of blocks \\	\hline
       $\gamma$  & Number of predicates\\	\hline	
       $\tau$ & Number of samples received \\ \hline
       $\kappa$ & An empirical constant to sequentially access one block \\ \hline
       $S = \{S_1,S_2,\cdots,S_\gamma\}$ & \densitymap indicated in the WHERE clause \\	\hline
       $S_{j}[i]$ & the $i$th entry of \densitymap $S_j$ \\	\hline
       $\hat{S} = \{\hat{S}_1,\hat{S}_2,\cdots,\hat{S}_\gamma\}$ & Sorted \densitymap indicated in the WHERE clause \\	\hline
       $\hat{S}_{i}[j]$ & the $j$th entry of the $i$th sorted \densitymap in $\hat{S}$ \\	\hline
       $\theta$ & Threshold \\ \hline
       $M$ & Set of block IDs with their aggregated densities\\ \hline
       $Seen$ & Set of block IDs seen so far \\ \hline
       $R$ & Set of block IDs returned by the algorithm
     \end{tabular}
 }
 \vspace{-10pt}\caption{Table of Notation}
  \label{tab:notation}
  \end{center}
\vspace{-25pt}
\end{table}

\subsection{{\large \thresh} Algorithm}
\label{subsec:thresh}
\thresh is based on the threshold algorithm proposed by Fagin et
al.~\cite{fagin2003optimal}.
The goal of \thresh 
is to use our in-memory \densitymap index to retrieve the densest blocks
until $k$ valid records are found.
The unmodified threshold algorithm by Fagin et al. would attempt to find the
$p$ densest blocks. However, in our setting, we do not know the value of $p$ in
advance:
we only know $k$, the number of valid tuples required, so we need to set the value
of $p$ on the fly. 

For fast execution of \thresh, an additional \emph{sorted \densitymap} data
structure is
required. For
every \densitymap $D$, we sort it in descending order of densities to create
a sorted \densitymap $\hat{D}$. Every element $\hat{D}[i]$ has two attributes: $bid$, the block ID,
and $density$, the percentage of tuples in this block which satisfies the
corresponding constraint. Here $D[1]$ refers to the first block of the
data and $\hat{D}[1]$ refers to the densest block in the data.
Sorted \densitymaps are precomputed during data loading time and stored in
memory along with the \densitymaps, so the sorting time does not affect the execution times of queries.

\stitle{High-Level Intuition}
At a high level, the algorithm examines each of the relevant sorted \densitymaps
corresponding to the predicates in the query.
It traverses these sorted \densitymaps in sorted order, while
maintaining a record of the blocks with the highest overall density for the query, 
i.e., the highest number of valid tuples.
The algorithm stops when the maintained blocks have at least $k$ valid records,
and 
it is guaranteed that none of the unexplored blocks
can have a higher overall density than the ones maintained.

\stitle{Algorithmic Details}
Algorithm~\ref{alg:threshold} provides the full pseudocode.
With sorted \densitymaps, it is easy to see how \thresh handles a query with
a single predicate: $A_i = V_i^j$. \thresh simply selects the $\hat{D}_i^j$ which
corresponds to the predicate and retrieves the first few blocks of $\hat{D}_i^j$ until
$k$ valid records are found.
For multiple predicates, the execution of \thresh is more complicated.
Depending on how the predicates are combined, $\bigoplus$ could mean $\prod$,
i.e., product, if the predicates are all combined using ANDs, or $\sum$, i.e.,
sum, if the predicates are all combined using ORs.
Each \densitymap in 
$\{S_1, ..., S_{\gamma}\}$ represents 
a predicate from the query, while $\{\hat{S}_1,...,\hat{S}_{\gamma}\}$
represent the sorted variants. 
At each iteration, we traverse down $\hat{S}_i$, while maintaining
a running threshold $\theta =
\bigoplus_{j=1}^{\gamma} \hat{S}_j[i].density$,
and also keeping track of all the block ids encountered across the sorted
density maps. 
This threshold $\theta$ represents the minimum aggregate density
that a block must have across the predicates before we are sure that it is one
of the densest blocks. 
During iteration $i$, we consider all blocks in $M$ examined in the previous iterations 
that have not yet been selected to be part of the output.
If the one with the highest density has density greater than $\theta$,
then it is added to the output $R$. 
We know that $\theta$ is an upper-bound for any blocks that have not already
been seen in this or the previous iterations, due to the monotonicity
of the operator $\bigoplus$. 
Thus, \thresh maintains the following invariant:
a block is selected to be part of the output iff its density is
as good or better than any of the blocks that not yet been selected to
be part of the output. 
Overall, \thresh ends up adding the blocks to the output in decreasing
order of density. 
\thresh terminates when the number of valid records in
the output blocks selected is at least $k$.

To retrieve the \anyk samples, 
we then load the blocks returned by \thresh into memory and return all valid
records seen in those blocks. If the total number of query results in those
blocks are less than $k$, we re-execute \thresh on the blocks that have not
been retrieved in previous invocations.

\noindent \emph{Fetch Optimization.} Depending on the order of the blocks returned by
\thresh, the system may perform many unnecessary random I/O operations. For
example, if \thresh returns blocks
%
$\{B_{100}, B_1, B_{83} , B_3\}$, the system may read block $B_{100}$, seek
to block $B_1$, and then seek back to block $B_{83}$, resulting in expensive
disk seeks. Instead, we can sort the blocks $\{B_1, B_3, B_{83}, B_{100}\}$ before
fetching them from disk, thereby minimizing random I/O and overall query execution
time.

{\small
\begin{algorithm}
\caption{\thresh}\label{alg:threshold}
\begin{algorithmic}[1]
\State Initialize $\theta \gets 0$, $i \gets 1$, $\tau \gets 0$, $R, M, Seen\gets \varnothing$

\While{$i \leq \lambda$}
	\State $\theta \gets \displaystyle \bigoplus_{j=1}^{\gamma}$ $\hat{S}_{j}[i].\textit{density}$
        \For {$j = 1  \dots  \gamma$}
		\If {$\hat{S}_j[i].bid \notin Seen$}
                \State $\rho \gets \hat{S}_j[i].bid$
                \State $\xi \gets \left\{
                  \begin{array}{ll}
                    bid: & \rho \\
                    density: & \bigoplus_{k=1}^{\gamma}
                    S_k[\rho].density
                  \end{array}
                  \right.
                  $
                \State $M \gets M \cup \{\xi \}$
			\State $Seen \gets Seen \cup \{\rho\}$
		\EndIf
	\EndFor
        \State $\mu \gets \argmax_{\mu' \in M} \; \mu'.density$
	\While{$\mu \geq \theta$ }
		\State $\tau \gets \tau + \mu.density \times records\_per\_block$
                \State $R \gets R \cup \{ \mu.bid \}$
                \State $M \gets M \setminus \{ \mu \}$
		\If {$\tau \geq k$} 
			\State \textbf{return} $R$
		\Else
                        \State $\mu \gets \argmax_{\mu' \in M} \; \mu'.density$
		\EndIf
	\EndWhile
	\State $i \gets i + 1$
\EndWhile
\State \textbf{return} $R$
\end{algorithmic}
\end{algorithm}
}

\stitle{Guarantees}
We now show that \thresh retrieves the minimum set of blocks
when optimizing for density.
\vspace{-3pt}
\begin{theorem}[Density Optimality]
Under the independence assumption,
\thresh returns the set of blocks 
with the highest densities with at least $k$ valid records.
\vspace{-3pt}
\end{theorem}
Since \thresh is a significant modification of the threshold algorithm
the proof of the above theorem does not follow directly from prior work.
\begin{proof}
The proof is composed of two parts:
first, we demonstrate that \thresh adds
blocks to $R$ in the order of decreasing overall density;
second, we demonstrate that \thresh stops only when the
number of valid records in $R$ is $\geq k$.
The second part is obvious from the pseudocode (line 16).
We focus on the first part.
The first part is proven using an inductive argument.
We assume that the blocks added to $R$ through $i$th iteration
satisfy the property and that $\theta$ of the $i$th iteration is denoted as
$\theta_i$.
We note that for the $i+1$th iteration,
$\theta_i \geq \theta_{i+1}$.
Consider the blocks that are part of $M$ at the end of line 10 in the $i+1$th
iteration.
These blocks fall into two categories:
either they were already part of $M$ in the $i$th iteration,
and hence have densities less than $\theta_i$,
or were added to $M$ in the $i+1$th iteration, 
and due to  monotonicity,  once again have density less
than $\theta_i$.
Furthermore, any blocks that have not yet been examined
will have densities less than $\theta_{i+1}$.
Since all blocks that have been added at iteration $i$ or prior
have densities greater than or equal to $\theta_i$,
all the blocks still under contention for adding to $R$---those in $M$
or those yet to be examined---have densities below those in $R$.
Now, in iteration $i+1$, we add all blocks in $M$ whose densities are
greater than $\theta_{i+1}$, in decreasing order. 
We know that all of these blocks have higher densities than all
the blocks that have yet to be examined (once again using monotonicity).
Thus, we have shown that any blocks added to $R$ in iteration $i+1$
are lower in terms of density than those added to $R$ previously, and are the best
among the ones in $M$ and those that will be encountered in future iterations.
\end{proof}

\subsection{{\large \prong} Algorithm}



Our second algorithm, \prong, prioritizes for locality rather than density,
aiming to identify the shortest sequence of blocks that
guarantee $k$ valid records.
The naive approach to identify this would be to consider
the sequence formed by every pair of blocks (along with 
all of the blocks in between)---leading to an algorithm that is quadratic in the 
number of blocks. 
Instead, \prong, described below, 
is linear
in the number of blocks. 

\stitle{High-level Intuition}
\prong moves across
the sequence of blocks using a sliding window formed using a start and an end pointer,
and eventually returns the smallest possible window with $k$ valid records.
At each point, \prong ensures that the window has at least $k$ valid
records within it, by first advancing the end pointer until we meet the constraint,
then advancing the start pointer until the constraint is once again violated.
It can be shown that this approach considers all minimal 
sequences of blocks with $k$ valid records. 
Subsequently, \prong returns the smallest such sequence. 

\stitle{Algorithmic Details}
\papertext{Additional details and pseudocode can be found in our technical report~\cite{needletail-tr}.}
\techreport{The pseudocode for the algorithm is listed in Algorithm~\ref{alg:prong}.}
The \prong algorithm operates on an array of values formed by 
applying the operator $\bigoplus$
to the predicate \densitymaps $\{S_1, .., S_{\gamma}\}$,
one block at a time. 
At the start, both pointers are at the value corresponding 
to the density of the 
first block.
We move the end pointer to the right until the number of 
valid records between the two pointers is no less than $k$;
at this point, we have our first candidate sequence 
containing at least $k$ valid records.
We then move the start pointer to the right, checking if each sequence
contains at least $k$ valid records, and continuing until the
constraint of having at least $k$ valid records is once again violated.
Afterwards, 
we once again operate on the end pointer. 
At all times, we maintain the smallest sequence found so far,
replacing it when we find a new sequence that is smaller.

\techreport{\small
\begin{algorithm}
\caption{\prong}\label{alg:prong}
\begin{algorithmic}[1]
\State Initialize $\tau \gets 0$, $R \gets \varnothing$
\State Initialize $start, end, min\_start, min\_end \gets 1 $
\For {$i = 1 \dots \lambda$}
    \State $M[i] \gets \left\{
      \begin{array}{ll}
        bid: & i \\
        density: & \bigoplus_{j=1}^{\gamma} S_{j}[i].density
      \end{array}
      \right.
      $
\EndFor

\While{$end < \lambda$}
	\While{$\tau < k$ and $end < \lambda$}
		\State $\tau$ $\gets \tau + M[end].density \times records\_per\_block$
		\State $end \gets end + 1$
	\EndWhile

	\While{$\tau \geq k$ and $start < \lambda$}
		\If {$(end - start) < (min\_end - min\_start) $}
			\State $min\_end \gets end$
			\State $min\_start \gets start$  
		\EndIf
		\State $\tau \gets \tau - M[start].density \times records\_per\_block$
		\State $start \gets start + 1$
	\EndWhile

\EndWhile
\State $R \gets \bigcup_{min\_start \leq i < min\_end} \{i\}$

\State \textbf{return} $R$
\end{algorithmic}
\end{algorithm}
}


%
%


\later{
\begin{example}[Two prong]
Considering the instance from \textit{Example 2}, \prong considers density maps $D_1^1 = \{(1,0.2),(2,0.1), $ $(3,0.3), (4, $ $ 0.4), (5,0.5), (6, 0.7), (7, 0.8), (8, 0.9), (9,0)\}$  and $D_2^2= \{(1, 0.1), $ $(2,0.3), (3, 0), (4, 0.9), (5, 0.6), (6, 0.7), (7, 0.1), (8, 0.8),($ $9, 0.2)\}$. We first calculate the resulting density map $M$, which is $\{(1, 0.02), (2, $ $0.03), (3, 0), (4, 0.36), (5, 0.3), (6, 0.49), (7, 0.08), (8, $ $0.72), (9, 0)\}$. Then \prong scans all entries in $M$ once and fetches blocks 6,7, and 8, as we estimated that this is the smallest contiguous set of blocks that contain at least 120 samples. 
\end{example}
}

\stitle{Guarantees}
We now show that \prong retrieves the minimum 
sequence of blocks when optimizing for locality.
\vspace{-3pt}
\begin{theorem}[Locality Optimality]
Under the independence assumption, \prong returns the smallest sequence of blocks that contains at least $k$ valid records. 
\end{theorem}
\vspace{-3pt}
\techreport{We demonstrate that for every block $i$, \prong considers the smallest sequence of blocks with $k$ valid
records beginning at block $i$ at some point in the algorithm, thereby proving the above theorem. }
\papertext{The proof can be found in the extended technical report~\cite{needletail-tr}.}
\techreport{\begin{proof}
For $i=1$, this is easy to see. The end pointer of \prong starts at 1 and
increases; the start pointer is not moved until a valid sequence of
blocks is found, so by construction \prong considers the smallest sequence of
blocks starting at 1.
For the remaining $i$'s we prove this by contradiction. Let the smallest
sequence of blocks
beginning at block $i$ end at $j$, where $j \ge i$; we denote this sequence as
$[i,j]$. Now, let $j'$ be the ending block for the smallest sequence of blocks
starting at $i+1$; this sequence is denoted as $[i+1,j']$. If $j' \ge j$, our
\prong algorithm considers the sequence as we move the end pointer forward
(lines 6-8 in the pseudocode). Assume to the contrary that $j' = j-1 < j$; that
is, the sequence $[i+1,j-1]$ is the smallest sequence of blocks starting at
$i+1$ that has $k$ valid records. $[i+1,j-1]$ is a subsequence of
$[i,j-1]$, so $[i,j-1]$ must also have at least $k$ valid records. However, we
already declared $[i,j]$ to be the smallest sequence of blocks starting at $i$
that has $k$ valid records, and thus a contradiction is found. Similar
arguments can be made for all $j' < j-1$, so \prong must consider the smallest
sequence of blocks starting at block $i$ for every $i$.
\end{proof}
}


\section{Hybrid ANY-K Algorithms}\label{sec:algo_hybrid}

The two desired properties of
density and locality
can often be at odds with each other
depending on the data layout; dense blocks may be far apart, and neighboring
blocks may contain many blocks which have no valid records. 
In this section, we first present a 
cost model to estimate
the I/O cost of a \anyk algorithm,
and use it to design an \anyk algorithm
that is I/O-optimal, providing
the best balance between density and locality,
and a hybrid algorithm,
that selects between \thresh and \prong. 





\subsection{A Simple I/O Cost Model}


\begin{figure}[tb]
 \centering
 \vspace{-10pt}
  \includegraphics[scale=0.55]{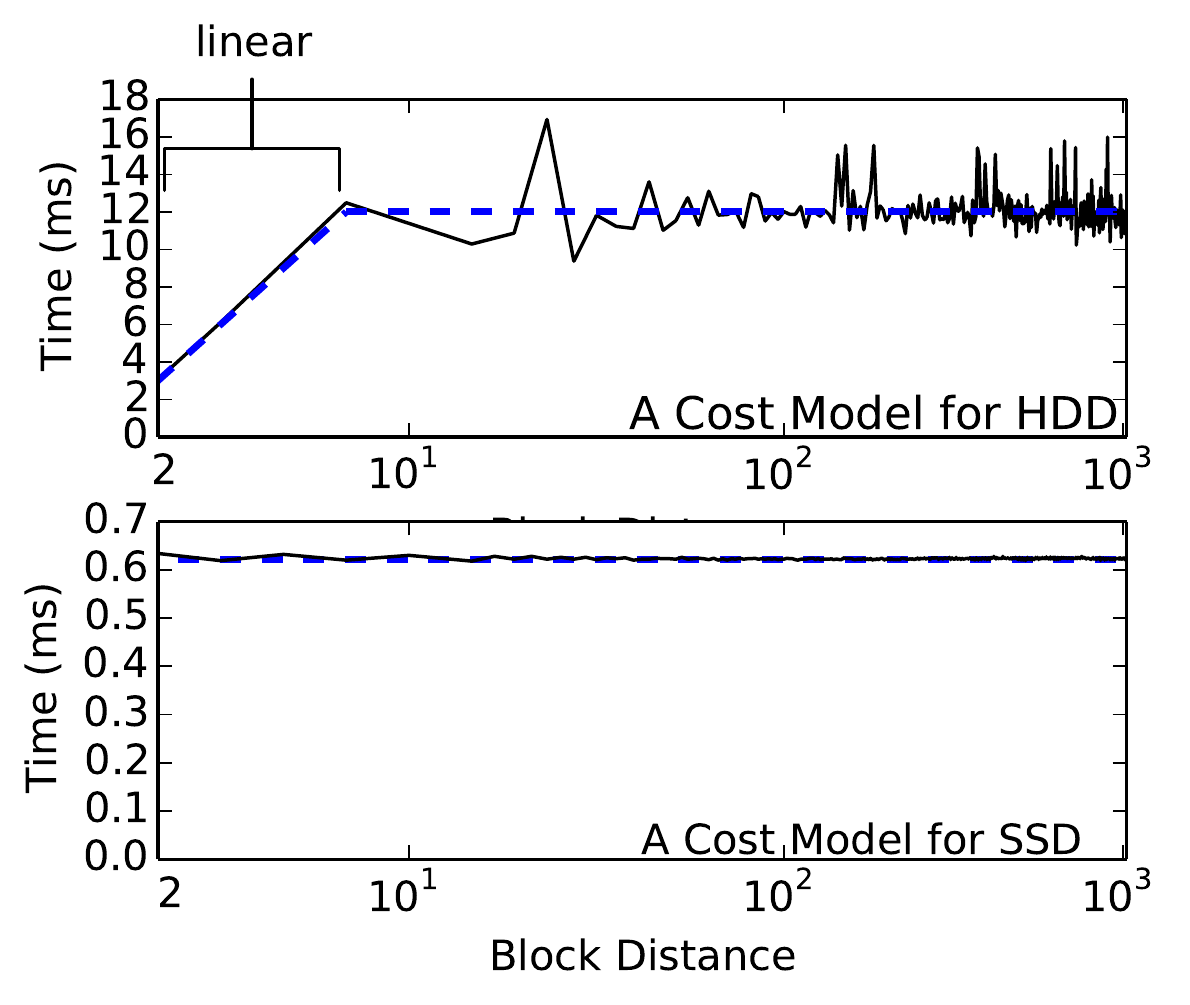}
  \vspace{-10pt}
  \caption{I/O Cost Model for HDDs and SSDs}
  \label{fig:cost_model}
\vspace{-10pt}
\end{figure}

To set up the cost model for I/O for HDDs (Hard Disk Drives), 
we profile the storage system as
described
by Ruemmler et al.~\cite{ruemmler1994introduction}. 
We randomly choose various starting blocks 
and record the time taken to fetch other blocks that are varying distances
away (where distance is measured in number of blocks),
for distance onwards. 
As shown in Figure~\ref{fig:cost_model},
which uses a linear scale on the x-axis from x=2 until x=10, and then
logarithmic scale after that, we
observe that with block size equal to 256KB, 
the I/O cost is smallest when doing a sequential I/O operation
to fetch the next block ($\sim$ 2ms), 
 and increases with the distance
up to a certain maximum distance $t$ after which it becomes constant
($\sim$ 12ms).
We have overlaid our cost model estimate using
a dashed blue line.
More formally, for two blocks $i$ and $j$, we model the cost of fetching
block $j$ after block $i$ as follows: 
\vspace{-3pt}
\[
  RandIO(i,j) = \left\{\def\arraystretch{1.2}%
  \begin{array}{@{}c@{\quad}l@{}}
    cost(i,j) & \text{if $\left|j-i\right| \le t$}\\
    constant & \text{$otherwise$}\\
  \end{array}\right.
  \vspace{-3pt}w
\]
When distance is less than $t$, we use a simple linear fit for $cost(i,j)$, using 
the Python \texttt{numpy.polyfit} 
function\footnote{\small
\url{https://docs.scipy.org/doc/numpy/reference/generated/numpy.polyfit.html}}

On the other hand, 
as shown in Figure~\ref{fig:cost_model}, the I/O Cost Model for SSDs 
is different from the one we see for HDDs.
Overall, we see a constant time ($\sim$ 0.6ms) to fetch a block
(overlaid in a dashed blue line)
independent of the block distance. 

\subsection{{\large \optimal} Algorithm}\label{sssec:dp}
\optimal considers both density and locality to search for the set of blocks
that will provide the minimum I/O cost overall. 
Specifically, given our cost model, we can use dynamic programming to
find the optimal set of blocks with $k$ valid records.

We define $C(s,i)$ as the minimal cost to retrieve $s$ estimated valid
records when block $i$ is amongst the blocks fetched. We define $Opt(s,i)$ as the
cost to retrieve the optimal set of blocks with $s$ estimated valid records
when considering the first $i$ blocks.
Finally, we denote $s_i$ as the estimated number of valid records inside block
$i$, derived, as before, using the $\bigoplus$ computation. 
With this notation, we have:
  \begin{align*}\label{eqn:DP}
  \small
    C(s,i) &= \min \left\{\begin{array}{lr}
        C(s-s_i, j) + RandIO(j,i), \quad \forall j \in [\,i-t, i-1]\,\\
        Opt (s-s_i, i-t-1) + RandIO(i-t-1,i) &
        \end{array}\right. \\
    Opt(s,i) &= \min \left\{\begin{array}{lr}
        C(s, i) \\
        Opt (s, i-1)
        \end{array}\right.
   \end{align*}
\noindent   
The intuition is as follows: for each block $i$ that has $s_i$ estimated valid records, either
the block can be in the final optimal set or not.
If we decide to include block $i$, the cost is the
minimum cost amongst the following:
\begin{inparaenum}[\itshape (i)\upshape]
\item
  the smallest I/O cost of having $s-s_i$ samples at block $j$ 
  where $|i-j| \le t$, plus the cost of jumping from block $j$ to $i$ (i.e.,
  $C(s-s_i,j) + RandIO(j,i)$), or
\item
the optimal cost at block $i-t-1$, plus the
random I/O cost of jumping from some block in the first $i-t-1$ blocks to block
$i$ (i.e., $Opt(s-s_i, i-t-1) + RandIO(i-t-1,i) )$.
\end{inparaenum}

For the second expression, 
if we exclude block $i$, then the optimal cost is the same as the optimal cost at block
$i-1$. Consequently, the optimal cost at block $i$ is the smallest value in
these two cases.
\papertext{The \optimal algorithm that implements this dynamic programming 
formulation can be found in the technical report~\cite{needletail-tr}.}
\techreport{The full algorithm is shown in Algorithm
~\ref{alg:optimal}, where $\kappa$ is some constant cost to fetch the first
block.}



\techreport{
{\small
\begin{algorithm}
\caption{\optimal}\label{alg:optimal}
\begin{algorithmic}[1]
\State Initialize $R \gets \varnothing$.
\For {$i = 1 \dots \lambda$}
    \State $M[i] \gets \left\{
      \begin{array}{ll}
        bid: & i \\
        density: & \bigoplus_{j=1}^{\gamma} S_{j}[i].density
      \end{array}
      \right.
      $
      \State $s_i$ $\gets$ $M[i].density \times records\_per\_block$
\EndFor

\For {$s$ = $0$ \dots $s_1$}
	\State $C(s, 1) \gets \kappa$
	\State $Opt(s, 1) \gets \kappa$
\EndFor
\For {$s$ = $s_1+1,$ \dots $,k$}
	\State $C(s, 1) \gets \infty$
	\State $Opt(s, 1) \gets \infty$
\EndFor

\For {$i$ = $2$ \dots $\lambda$}
	\For {$s$ = $0$ \dots $s_i$}
		\State $C(s,i) \gets RandIO(1,i)$
		\State $Opt(s,i) \gets \min{\{Opt(s, i-1), C(s,i)\}}$
	\EndFor
	\For{$j$ = $s_i+1,$ \dots $,k$}
		\State {\small $C(s,i) \gets \min \left\{\begin{array}{lr}
     							C(s-s_i, j) +
                                                        RandIO(j,i), \forall j \in [\,i-t, i-1]\,\\
   						        Opt (s-s_i, i-t-1) +
                                                        RandIO(i-t-1, i) &
     				  \end{array}\right.$
              }
		\State $Opt(s,i) \gets \min{\{Opt(s, i-1), C(s,i)\}}$
	\EndFor
\EndFor

\State $R \gets$ sequence of blocks that result the cost in $Opt(k, \lambda)$
\State \textbf{return} $R$
\end{algorithmic}
\end{algorithm}
}
}

\stitle{Guarantees}
We can show the following property.
\vspace{-3pt}
\begin{theorem}[\optimal]
Under the independence assumption and the constructed cost model for disk I/O, 
\optimal gives the blocks with optimal I/O cost for fetching \anyk valid records. 
\end{theorem}
\vspace{-3pt}
The proof is listed in full detail in
\techreport{Appendix~\ref{sec:appendix-proof}.} 
\papertext{the technical report~\cite{needletail-tr}.}


\later{
\subsection{Variants of {\large \thresh}}
\label{subsec:variants}
\subsubsection{Arbitrary Nestings of ANDs and ORs}
In the previous sections, we assumed that query predicates were combined
with either all ANDs or all ORs, thus leading $\bigoplus$ to be either $\prod$
or $\sum$. However, real-world queries feature complex query predicates which
can have arbitrary nestings of ANDs and ORs. To address this, we can extend the
definition of $\bigoplus$. Assuming $\bigoplus(P)$
calculates the density of complex predicate $P$, we can define $\bigoplus$
recursively:
\[
  \bigoplus(P) = \left\{
  \begin{array}{ll}
    \bigoplus(A) \bigoplus(B) & \text{if} \quad P = A\text{ AND }B \\
    \bigoplus(A) + \bigoplus(B) & \text{if} \quad P = A\text{ OR }B
  \end{array}
  \right.
\]

\subsubsection{Changing the Granularity}
Although \optimal strikes the optimal balance between density and locality, we
see later in Section~\ref{subsec:opt-eval} that it can be computationally expensive to
perform. To address this, we heuristically consider groups of consecutive blocks (``superblocks'')  in\optimal, which will trade off density for locality. Using bigger superblocks yields
more locality as each superblock now encompasses more blocks.
We explore the effects of changing superblock sizes (\emph{granularity}) in
Section~\ref{subsec:params}.
}

\subsection{{\large \hybridactual} Algorithm}
\label{subsec:hybridactual}
Even though \optimal is able to return the optimal I/O cost 
for fetching \anyk samples, its much higher computation cost 
(as we show in our experiments) makes it impractical for large datasets. 
We propose \hybridactual
which simply selects between the best of \thresh
and \prong, using our I/O cost model, when a query is issued.
Since \hybridactual needs to run both algorithms 
to determine the set of blocks selected
by each algorithm, using \hybridactual would involve
a higher up-front computational cost,
but as we will see, leads to substantial performance benefits.




\section{Aggregate Estimation}
\label{sec:est}
So far, our \anyk algorithms retrieve 
$k$ records 
without any consideration of how representative they are of the entire population
of valid records.
If these records are used to estimate an aggregate 
(e.g., a mean), there could be bias in this value due
to possible correlations between the value and the data layout.
While this is fine for browsing, it leaves the user unable to make any
statistically significant claims about the aggregated value. 
To address this problem, we make two simple adjustments
to extend our \anyk algorithms. First, we introduce
a \emph{\hybrid sampling scheme} 
where we add small amounts of random data to our \anyk estimates. 
\techreport{This random data is added in a fashion such that it does not significantly affect
the overall running time, 
while at the same time, allowing us to 
``correct for'' the bias easily.}
Second, we correct the bias
by leveraging  techniques from the
survey sampling literature.\techreport{ Specifically, we leverage the
Horvitz-Thompson~\cite{horvitz1952generalization} and
ratio~\cite{lohr2009sampling} estimators, as described below.}
Note that such approaches have been employed in other settings
in approximate query processing~\cite{mozafari2015handbook,li2016wander,cohen2011structure,hu2009estimating, kandula2016quickr}, 
but their application to \anyk is new. 


%
%

\subsection{{\large \hybrid} Sampling}
We propose a \hybrid sampling scheme, in which we collect a large
portion of the $k$ requested samples using an \anyk algorithm, and collect the
rest in a random fashion.
We denote $(1-\alpha)$ as the proportion of $k$ samples we retrieve using
the \anyk algorithm,
and $\alpha$ as the proportion of $k$ samples we retrieve using random sampling.
The user chooses the parameter $\alpha$ upfront based on how much random
sampling they wish to add. 
While a larger
$\alpha$ may reduce the number of total samples needed to obtain a statistically
significant result, the time taken to retrieve random samples greatly exceeds
the time taken to retrieve samples based on our \anyk algorithms.
Therefore, $\alpha$ needs to be carefully chosen;
we experiment with different $\alpha$s in
Section~\ref{sec:eval}.
\papertext{We provide a more formal description of our sampling
procedure in our technical report~\cite{needletail-tr}.}

\techreport{
More formally, if we let $S_v$ be the set of
blocks which have at least one valid record in them,
we can describe
\hybrid sampling as follows:
\begin{inparaenum}[(1)]
\item
  Use an \anyk algorithm to choose the densest blocks $S_c$ from $S_v$, and derive
  $(1-\alpha) k$ samples from $S_c$.
\item
  Uniformly randomly select blocks $S_r$ from the remaining blocks, and derive
  $\alpha$ samples from $S_r$.
\end{inparaenum}
Note that $S_c \cap S_r = \varnothing$.
}




 
\subsection{Unequal Probability Estimation}
Within the {\hybrid sampling scheme}, 
the probability a block is sampled is not
uniform.
Therefore,
we must use an \emph{unequal probability estimator}~\cite{thompson2012sampling} and 
inversely weigh samples based
on their selection probabilities.
\papertext{We apply two different estimators for this: 
the Horvitz-Thompson~\cite{horvitz1952generalization} and
ratio~\cite{lohr2009sampling} estimators.
Even though the Horvitz-Thompson estimator is an unbiased estimator, 
its variance is known to often be large. 
On the other hand, the ratio estimator
is known to have low variance.
As shown later in Section~\ref{subsec:time_err}, the ratio
estimator works quite well in situations where aggregate estimate is not
correlated with the block densities. 
We describe further details and formulae in our technical report~\cite{needletail-tr}.}
\techreport{We introduce two different estimators for this: the
Horvitz-Thompson estimator and the ratio estimator.}

%
\techreport{
\subsubsection{Requisite Notation}
The goal of our two estimators is to estimate the true aggregate sum $\tau$ and the true
aggregate mean $\mu$ of measure attribute $M$ given a query $Q$. 
We use $\tau_i$ to denote the aggregate sum of $M$ for block $i$ and $L$ for the
total number
of valid records for query $Q$. We can estimate $L$ using the \densitymaps.

The estimators inversely weigh samples based on their probability of
selection. So, we define the \emph{inclusion probability} $\pi_i$ as the probability
that block $i$ is included in the overall sample:
\[
  \pi_i =
  \begin{cases}
    1 & \text{if } i \in S_c \\
    \frac{|S_r|}{|S_v| - |S_c|} & \text{if } i \in S_v \setminus S_c \\
    0 & \text{otherwise}
  \end{cases}
\]

For the $(1-\alpha)k$ samples that come from the \anyk blocks in $S_c$, the
probability of being chosen is always 1. After these blocks have been selected,
a uniformly random subset of the remaining blocks are chosen to
produce the $\alpha k$ random samples; thus the probability that these samples
are chosen is
$\frac{|S_r|}{|S_v|-|S_c|}$.

%


We define the \emph{joint inclusion probability} $\pi_{ij}$ as the probability
of selecting both blocks $i$ and $j$ for the overall sample:
\[
  \pi_{ij} = 
  \begin{cases}
    1 & \text{if } i \in S_c \land j \in S_c \\
    \frac{|S_r|}{|S_v| - |S_c|} & \text{if } (i \in S_c \land j \in S_r) \lor
    (i \in S_r \land j \in S_c) \\
    \frac{|S_r|}{|S_v|-|S_c|} \frac{|S_r|-1}{|S_v|-|S_c|-1} & \text{if } i \in
    S_r \land j \in S_r \\
    0 & \text{otherwise}
  \end{cases}
\]




\subsubsection{Horvitz-Thompson Estimator}
%
Using the Horvitz-Thompson~\cite{horvitz1952generalization} estimator, $\tau$ is estimated as:
\begin{equation}\label{est_total}
  \hat{\tau}_{HT} = \sum_{i \in S_c} \frac{\tau_i}{\pi_i} + \sum_{i \in S_r} \frac{\tau_i}{\pi_i}
\end{equation}

As mentioned before, the sums $\tau_i$ are inversely weighted
by their probabilities $\pi_i$ to account for the different probabilities of
selecting blocks in $S_v$.
Based on $\hat{\tau}_{HT}$, we can also easily estimate $\mu$ by dividing the
size of the population:
\begin{equation}
  \hat{\mu}_{HT} = \frac{\hat{\tau}_{HT}}{L}
\end{equation}

The Horvitz-Thompson estimator guarantees us that both $\hat{\tau}_{HT}$ and
$\hat{\mu}_{HT}$ are unbiased estimates: $E(\hat{\tau}_{HT}) =
\tau$ and $E(\hat{\mu}_{HT}) = \mu$. A full proof can be found in~\cite{horvitz1952generalization}.
In addition, the Horvitz-Thompson estimator gives us a way to calculate the variances of
of $\hat{\tau}_{HT}$ and $\hat{\mu}_{HT}$, which represent the expected bounds
of $\hat{\tau}_{HT}$ and $\hat{\mu}_{HT}$:
%
%
\begin{equation}
  Var({\hat{\tau}_{HT}}) =
\sum_{i \in S_v} \left(\frac{1-\pi_i}{\pi_i} \right) \tau_i^2 + \sum_{i \in
S_v}\sum_{j \neq i} \left(\frac{\pi_{ij} - \pi_i \pi_j}{\pi_i \pi_j}\right)
\tau_i \tau_j
\label{eq:var-tau-ht}
\end{equation}
\begin{equation}
  Var(\hat{\mu}_{HT}) = Var (\hat{\tau}_{HT} /L) = Var(\hat{\tau}_{HT}) /L^2
\label{eq:var-mu-ht}
\end{equation}


\subsubsection{Ratio Estimator}
Although the Horvitz-Thompson estimator is an unbiased estimator, it is
possible that the variances given by Equations \ref{eq:var-tau-ht} and
\ref{eq:var-mu-ht} can be quite large if the
aggregated variable is not well related to the inclusion
probabilities~\cite{thompson2012sampling}.
To reduce the variance,
the ratio
estimator~\cite{lohr2009sampling} may be used:
\begin{equation}
  \hat{\mu}_R = \frac{\hat{\tau}_{HT}}{\sum_{i \in S_c \cup S_r}
  \frac{L_i}{\pi_i}}
  \label{eq:mu-r}
\end{equation}
\begin{equation}
  \hat{\tau}_R = \hat{\mu}_R L
\end{equation}
where $L_i$ is the number of valid records in block $i$. The variances of
$\hat{\mu}_R$ and $\hat{\tau}_R$ are given by:
\begin{align}
  Var(\hat{\mu}_R) = &\frac{1}{L^2} \Bigg[ \sum_{i \in S_v} \left(\frac{1 -
      \pi_i}{\pi_i}\right) (\tau_i - \mu)^2 \nonumber \\
      &+ \sum_{i \in S_v} \sum_{j \ne i}
      \left(\frac{\pi_{ij} - \pi_i \pi_j}{\pi_i \pi_j}\right) (\tau_i -
    \mu)(\tau_j - \mu) \Bigg]
\end{align}
\begin{equation}
  Var(\hat{\tau}_R) = Var(\hat{\mu}_R L) = L^2 Var(\hat{\mu}_R)
\end{equation}

While the ratio
estimator is not precisely unbiased, in Equation~\ref{eq:mu-r},
we see that the numerator is the unbiased Horvitz-Thompson estimate of the sum
and the denominator is an unbiased Horvitz-Thompson estimate of the total
number of valid records, so the bias tends to be small and decreases with
increasing sample size.
}

We compare the empirical accuracies of these
two estimators in Section~\ref{sec:eval},
and demonstrate how our {\hybrid sampling technique}, when employing these
estimators,
provides accurate estimates of various aggregates values.

\section{Grouping and Joins}
\label{sec:join}

So far, we have assumed that all our sampling queries have the form
dictated by the SELECT query given in Section~\ref{sec:prob}, thus limiting our
operations to
a single database table, with simple selection predicates
and no group-by operators.
We now extend the \anyk sampling problem formulation and our
algorithms to handle more complex queries
that involve grouping and join operations.

\subsection{Supporting Grouping}

Rather than computing a simple \anyk, users may want to retrieve $k$ values per group, e.g.,
to compute an estimate of an aggregate value in each group.

Although a trivial way to do this would be to run a separate \anyk query per group, in this section
we discuss an algorithm that can share the computation across groups in the common case when 
users want $k$ values per group.

Consider an \anyk query $Q_k$ over a table $T$ with $S$ representing
the predicate in the where clause. Let $A_G$ be the grouping attribute with
values in $\{V_G^1, V_G^2, \ldots,V_G^{\delta_G}\}$.
The formal goal of this  grouped \anyk sampling can be stated as:
\vspace{-3pt}
\begin{problem}[grouped \anyk sampling]
  Given a query $Q_k$ defined by a predicate S on table $T$,
  and a grouping attribute $A_G$, the goal of grouped
  \anyk sampling is to retrieve any $k$ valid records for each group in as
  little time as possible.
\end{problem}
\vspace{-3pt}


Our basic approach is to
create a combined density map, which takes into account every group in the
group-by operation, and run the \anyk algorithm for all groups at once. This is
akin to sharing scans in
traditional databases.

In order to run our \anyk algorithms for all groups, we first define the
\emph{combined density} of the $l$th block as multiplication of two factors:
(1) the density of the $l$th block with respect to predicate $S$, 
and (2) the sum of the densities for group values in $A_G$ in the
$l$th block which still need to be sampled.
The first factor has been discussed previously, while
the second factor can be defined as:
\begin{equation}
\vspace{-3pt}
  d_{G_l}^* = \frac{1}{\text{RPB}}\sum_{j=1}^{\delta_G} min\left(k - r_G^j\;, \; d_{G_l}^j
  \times \text{RPB} \right)
\vspace{-3pt}%
  \label{eq:gby-opt-heuristic}
\end{equation}
where RPB is $records\_per\_block$, $r_G^j$
is the number of samples already
retrieved for group $V_G^j$,
 and
$d_{G_l}^j$ is the density of the $l$th block for the value $V_G^j$.
The expression inside the $min$ function
estimates the number of expected records in block $l$ for each group
$V_G^j$, but limits the estimate by the number of samples left to be retrieved
for that group\footnote{\small $r_G^j$ is never greater than $k$, so 
the $k - r_G^j$ expression cannot be negative.}. Thus, the combined density ($d_{S_l}
d_{G_l}^*$), 
where $d_{S_l}$ is the density of the $l$th block with respect to
predicate $S$, gives priority
to groups which have had fewer than k samples retrieved so far, and groups
which already have $k$ samples no longer contribute to the combined density.
The $1/$RPB in front of the summation for $d_{G_l}^*$ acts a normalization factor to ensure
  that $d_{G_l}^*$, and thereby $d_{S_l} d_{G_l}^*$, are both density values between 0 and
1.


With this combined density estimate $d_{S_l} d_{G_l}^*$, we can now construct an iterative
\anyk algorithm for grouped sampling operations, similar to the algorithms in
Sections~\ref{sec:alg} and~\ref{sec:algo_hybrid}. The main structure of the algorithm is as follows:
\begin{inparaenum}[(1)]
\item
  Update all densities using with $d_{S_l} d_{G_l}^*$.
\item
  Run one of the \anyk algorithms to retrieve $\psi$ blocks with the highest
  combined density.
\item
  Update the densities of the $\psi$ blocks as 0.
\item
  If $k$ samples still have not been retrieved for each group, go back to step (1).
\end{inparaenum}
Since $d_{G_l}^*$ depends on the number of samples already
retrieved, it must be updated periodically to ensure the correctness
of the combined densities. The $\psi$ parameter controls the periodicity of these
updates. The problem of setting $\psi$ becomes a trade-off between CPU time and
I/O time.
Setting $\psi=1$ updates the densities after every block retrieval; while this
more correctly prioritizes blocks and is likely to lead to fewer blocks
retrieved overall, there is a high CPU cost in updating the densities after each
block retrieval. As $\psi$ increases, the CPU cost goes down due to less frequent
updates, but the overall I/O cost is likely to go up since the combined
densities are
not completely up-to-date for each block retrieved.
Although our iterative algorithm is not particularly complex, and globally IO-optimal
solutions may perform better than our locally optimal solution, our algorithm
has the advantage of simplicity of implementation and likely lower CPU
overhead.
We defer consideration of more sophisticated algorithms to future work.
\papertext{We present the full pseudocode for our iterative algorithm and extensions to
multi-attribute grouping in the technical report~\cite{needletail-tr}.}

\techreport{
\stitle{Algorithm Details}
The full algorithm for the grouping \anyk query is shown in
Algorithm~\ref{alg:group-by}. In Section~\ref{sec:alg}, $\tau$ was a single
value representing the number of samples retrieved; for the grouping \anyk
algorithm, $\tau$ is now an array of size $\delta_G$ where each entry
represents the number of samples for that group.
Every iteration
consists of updating the combined density estimates or the \emph{priorities} of blocks $M$ based on the number of samples
retrieved (setting it to 0 if it has already been seen), and calling an \anyk
algorithm with $M$ and the number of blocks desired $\psi$. The algorithm
updates the counts of $\tau$ and the algorithm only ends once every entry in
$\tau$ is at least $k$.

{\small
\begin{algorithm}
\caption{Group-by \anyk algorithm.}\label{alg:group-by}
\begin{algorithmic}[1]
  \State Initialize $\tau \gets [0,...,0]$, $R, M \gets \varnothing$

\While{$\exists j \in \{1,...,\delta_G\}, \; \tau[j] < k$}
\For {$i = 1 \dots \lambda$}
\If{$i \in R$}
\State $M[i] \gets \left\{
  \begin{array}{ll}
    bid: & i \\
    density: & 0 
  \end{array}
  \right.$
  \Else
  \State $M[i] \gets \left\{
  \begin{array}{ll}
    bid: & i \\
    density: & d_{S_i} \sum_{j=1}^{\delta_G} d_{G_i}^*
  \end{array}
  \right.$
  \EndIf
  \EndFor
  \State $R' \gets \text{\anyk}(M, \psi)$
  \For{$r \in R'$}
  \For{$j \in \{1,...,\delta_G\}$}
  \State $\tau[j] \gets \tau[j] + d_{S_r} d_{G_r}^j \times records\_per\_block$
  \EndFor
  \EndFor
  \State $R \gets R \cup R'$
\EndWhile
\State \textbf{return} $R$
\end{algorithmic}
\end{algorithm}
}}

As we show in the next section, the key-foreign key join \anyk problem is
essentially equivalent to this grouped \anyk formulation, and we evaluate the
performance of our algorithm on these problems in
Section~\ref{subsec:join_exp}.

\techreport{
\stitle{Optimal Solution}
As mentioned, the grouping \anyk solution uses a heuristic to find the
best blocks to retrieve. However, an I/O optimal solution, similar to \optimal from
Section~\ref{sssec:dp}, could be derived using dynamic programming with 
a recursive relationship based on the notion of priority from
Equation~\ref{eq:gby-opt-heuristic}, $\tau$, and the disk model from
Section~\ref{sssec:dp}. Unfortunately, the resulting dynamic programming
solution becomes a complex program of even more dimensions
than the program from Section~\ref{sssec:dp}. Since we
already showed in Section~\ref{sec:eval} that \optimal incurs a prohibitively
high CPU cost in exchange for its optimal I/O time, we chose not to pursue this
avenue.

\stitle{Multiple Groupings} For multiple group-by attributes, we simply extend
the above formula to account for every possible combination of values from the
different groupings. For example, if we have two group-by attributes $A_G$ and
$A_{G'}$ \prev{with \densitymaps $\{D_G^1, ..., D_G^{\delta_G}\}$ and $\{D_{G'}^1,
  ..., D_{G'}^{\delta_{G'}}\}$ respectively}, 
we can specify our updated notion
 of density with  $d_{S_l} \sum_{j=1}^{\delta_G}
  \sum_{i=1}^{\delta_{G'}} d_{G_l}^* d_{{G'}_l}^* .$


}

\subsection{Supporting Key-Foreign Key Joins}

Consider \anyk sampling on the result of 
a key-foreign key join between two 
tables $T$ and $T'$,
where $A_J$ is the primary key in $T$, and $A_{J'}$ is the foreign
key in $T'$.
Similar to grouping, the formal definition of join \anyk sampling
can be defined as:
\vspace{-3pt}
\begin{problem}[join \anyk sampling]
  Given a query $Q_k$ defined by a predicate $S$ and a join
  over tables $T$ and $T'$, on primary key $A_J$ from table $T$ and foreign key $A_{J'}$
  from table $T'$, the goal of join
  \anyk sampling is to retrieve any $k$ valid joined records for each join value in as
  little time as possible.
\end{problem}
\vspace{-3pt}
For example,
if we want to join on a ``{departments}'' attribute, $k$ samples would be
retrieved for each department.

Since we assume $A_J$ is the primary key, and therefore unique, the
join \anyk sampling problem can be reduced to finding any $k$ valid records in
table $T'$ for each join value $V_J \in A_J$. However, this is the exact same
problem as the grouped \anyk sampling problem in which the group values are
$V_J \in A_J$.  Thus, we can use the algorithm described in the previous section,
using the values of $A_J$ as the grouping value on the foreign key table $T'$.


In this way, \ntail is able to best indicate the blocks that can be retrieved to minimize
the overall time for joins. We evaluate our join
\anyk algorithm in Section~\ref{subsec:join_exp}. We leave optimizations for other
join variants as future work.

\section{System Design}


\label{sec:arch}
\begin{figure}[htb]
\vspace{-5pt}
\centering
\includegraphics[width=.8\linewidth]{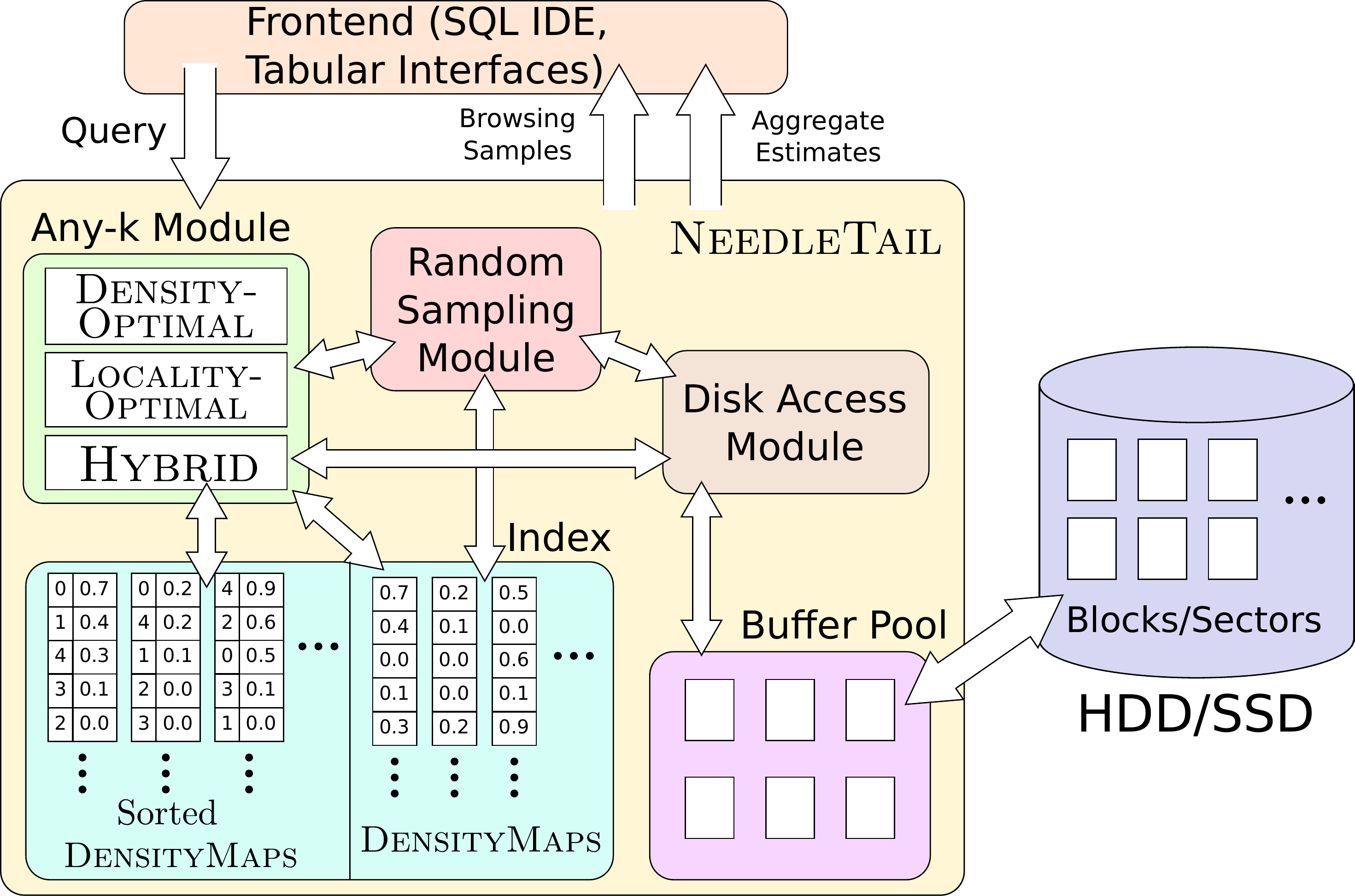}
\vspace{-3pt}
\caption{\ntail Architecture}
\label{fig:arch}
\vspace{-8pt}
\end{figure}

We implemented our \densitymaps, \anyk algorithms, and aggregate estimators in
a system called \ntail. 
\ntail is developed as a standalone browsing-based data exploration engine, 
capable of returning individual records as well as estimating aggregates. 
\ntail can be invoked by various frontends, e.g., SQL IDEs or interfaces
such as Tableau or Excel. 
Figure~\ref{fig:arch} depicts the overall architecture of
\ntail. It includes four major components: the \anyk module, the random
sampling module, the index, and the disk access module.  The \anyk module
receives queries from the user and executes our \anyk algorithms from
Sections~\ref{sec:alg}, \ref{sec:algo_hybrid}, and \ref{sec:join}
 to return \anyk browsing samples as quickly as possible.
For aggregate queries, the random sampling module is used in conjunction with
the \anyk module to perform the \hybrid sampling from
Section~\ref{sec:est}. The index contains the \densitymaps and sorted
\densitymaps.  Finally,
the disk access module is in charge of interacting with the buffer pool to
retrieve blocks from disk. Since \densitymaps are a lossy compression of the
original bitmaps, it is possible that some blocks with no valid records may be
returned; these blocks are filtered out by the disk access module.





%
%
%
%

Our \ntail prototype is currently implemented in C++ using about 5000 lines of
code. It is capable of reading in row-oriented databases with 
\texttt{int}, \texttt{float}, and \texttt{varchar} types and supports
Boolean-logic predicates. 
\techreport{Although the current implementation is limited to
a single machine, we plan to extend \ntail to run in a distributed environment
in the future. We believe the collective memory space available in a
distributed environment will allow us to leverage the \densitymaps in even
better ways.}

%
%
%


\section{Performance Evaluation}
\label{sec:eval}
In this section, we evaluate \ntail, focusing on 
runtime, memory consumption, and accuracy of estimates. 
We show that our \densitymap-based
\anyk algorithms
outperform any ``first-to-$k$-samples'' algorithms using
traditional OLAP indexing structures such as bitmaps or compressed bitmaps on a
variety of synthetic and real datasets. In addition, we empirically demonstrate that our
\hybrid sampling scheme is capable of achieving as accurate an aggregate estimation
as random sampling in a fraction of the time. 
Then, we demonstrate that our join \anyk algorithms provide substantial
speedups for key-foreign key joins.\techreport{ We conclude the section with an
exploration into the effects of different parameters on our \anyk
algorithms.}\papertext{ In our technical report~\cite{needletail-tr}, we study the impact of a 
number of parameters on \ntail including:
\begin{inparaenum}[\itshape (i)\upshape]
\item data size,
\item number of predicates,
\item density,
\item block size, and
\item granularity.
\end{inparaenum}}

%
%
%
%

\subsection{Experimental Settings}
We now describe our experimental workload, the evaluated 
algorithms, and the experimental setup.

\noindent\textbf{Synthetic Workload}: We generated 10 clustered synthetic datasets using
the data generation model described by Anh and
Moffat~\cite{clustereddatagenerator}. Every synthetic dataset has 100 million
records, 8 dimension attributes, and 2 measure attributes. For the sake of
simplicity, we forced every dimension attribute to be binary (i.e., valid
values were either 0 or 1), and with measure attributes being sampled from normal distributions,
independent of the dimension attributes. 
For each dimension attribute, we enforced an overall density of
10\%; the number of 1's for any attribute was 10\% of the overall number of
records.  Since we randomly generated the clusters of 1's in each attribute value,  
we ran queries with equality-based predicates on the first two dimensional attributes (i.e., $A_1=0$ and $A_2=1$).
Note that this does not always result in a selectivity of 10\% since
the records whose $A_1=0$ may not have $A_2=1$.

\noindent \textbf{Real Workload}: We also used two real datasets.
\later{
Since the performance is impacted by the layout of the data,
we also report how the data is laid out in the real dataset;
we preserve this layout in our experiments. 
}
\begin{denselist}

  \item \textbf{Airline Dataset~\cite{airlinedata}}: This dataset contained the
    details of all flights within the USA from 1987--2008, sorted based on time.
     It consisted of 123 million rows and 11 attributes with a total size of 11 GB.  
     We ran 5 queries with 1 to 3 predicates on attributes such as origin airport,
    destination airport, flight-carrier, month, day of week.
    For our experiments on error (described later),
    we estimated the average arrival delay, average departure delay, and average
    elapsed time for flights. 

  \item \textbf{NYC Taxi Dataset~\cite{nyctaxidata}}: This dataset contained
    logs for a variety of taxi companies in New York City for the years 2014 and 2015.
    The dataset as provided was first sorted by the year; within each year, it was first 
    sorted on the three taxi types and then on time. It
    consisted of 253 million rows and 11 attributes with a total size of 21
    GB. We ran 5 queries with 1 to 2 predicates on attributes including pickup
    location, dropoff location, time slots, month, passenger count, vendors,
    and taxi type. 
    For our experiments on error, we estimated the
    average fare amount and average distance traveled for the chosen trips.

\end{denselist} 


\noindent \textbf{Algorithms}:
We evaluated the performance of the three \anyk algorithms presented in
Section~\ref{sec:alg} and \ref{sec:algo_hybrid}:
\begin{inparaenum}[\itshape (i)\upshape]
\item \optimal,
\item \thresh,
\item \prong, and
\item \hybridactual

\end{inparaenum} We compared our algorithms against the following four
``first-to-$k$-samples'' baselines. \bmscan and \scan are representative of how
current databases implement the LIMIT clause.

%
%
\begin{denselist}
\item \bmscan:
  Assuming we have bitmaps for every predicate, we use bitwise AND and OR
  operations to construct a resultant bitmap corresponding to the valid
  records. We then retrieve the first $k$ records whose bits are set in this bitmap.


\item \lossy \cite{wiki:lossybitmap}:
\lossy is a variant of bitmap indexes where a bit is set for each block
instead of each record. For each attribute value, a set bit for a block
indicates that at least one record in that block has that attribute value.
During data retrieval, we perform bitwise AND or OR operations and on these
bitmaps then fetch
$k$ records from the first few blocks which their bit set.
\techreport{
Note that this is equivalent to a \densitymap which rounds its 
densities up to 1 if it is $>0$.}

\item \ewah: 
This baseline is identical to \bmscan, except the bitmaps are compressed using
the Enhanced Word-Aligned Hybrid (EWAH) technique~\cite{lemire2010sorting}
implemented using ~\cite{ewah_impl}

%

\item \scan:
Without using any index structures, we continuously scan the data on
disk until we retrieve $k$ valid records. 
%
\end{denselist}
For our experiments on aggregate estimation,
we compared our \hybrid sampling algorithms against the baseline \bmrand,
which is similar to \bmscan, except that it selects
$k$ random records among all the valid records.
We describe our setup for the join \anyk  experiments in Section~\ref{subsec:join_exp}.

\begin{figure}[tb]
 \centering
  \includegraphics[scale=0.5]{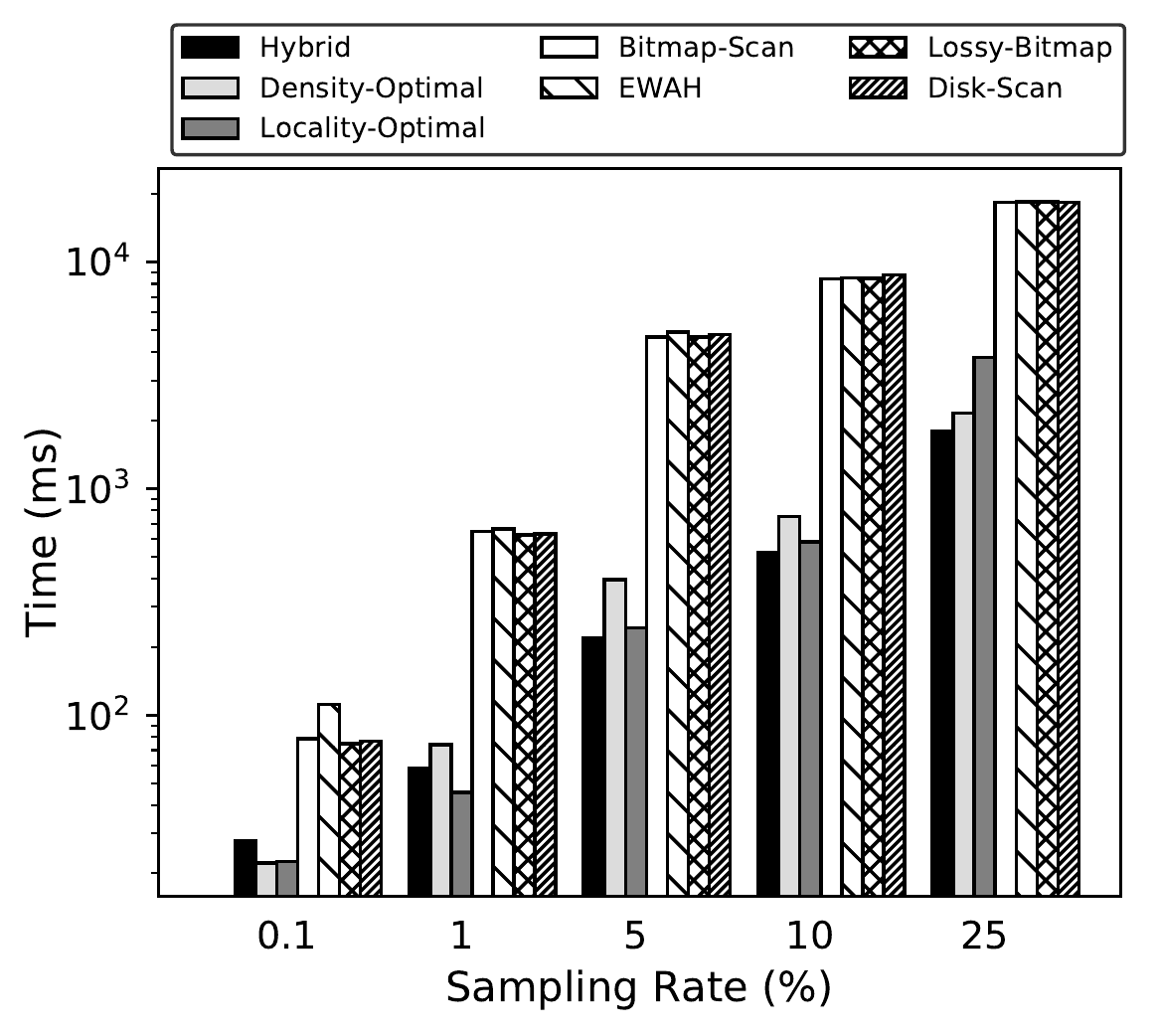}
  \vspace{-10pt}
  \caption{Query runtimes for the synthetic workload on a HDD.}
  \label{fig:syndata}
\vspace{-10pt}
\end{figure}

\noindent \textbf{Setup}:
All experiments were conducted on a 64-bit Linux server with 8 3.40GHz Intel
Xeon E3-1240 4-core processors and 8GB of 1600 MHz DDR3 main memory.
We tested our algorithms with a 7200rpm 1TB HDD and a 350GB SSD.
For each experimental
setting, we ran 5 trials (30 trials for the random sampling experiments) for
each query on each dataset. In every trial, we measured the end-to-end runtime, the CPU
time, the I/O time, and the memory consumption. Before each
trial, we dropped the operating system page cache and filled it with dummy
blocks to ensure the algorithms did not leverage any hidden benefits from the
page cache. To minimize experimental variance, we discarded the trials with the maximum and minimum
runtime and reported the average of the remaining.
Finally, after empirically testing a few different block sizes, we
found 256KB to be a good default block size for our datasets:
the block size does not significantly impact the relative performance 
of the algorithms.


\begin{figure*}[tb]
  \centering
  \vspace{-20pt}
  \includegraphics[scale=0.5]{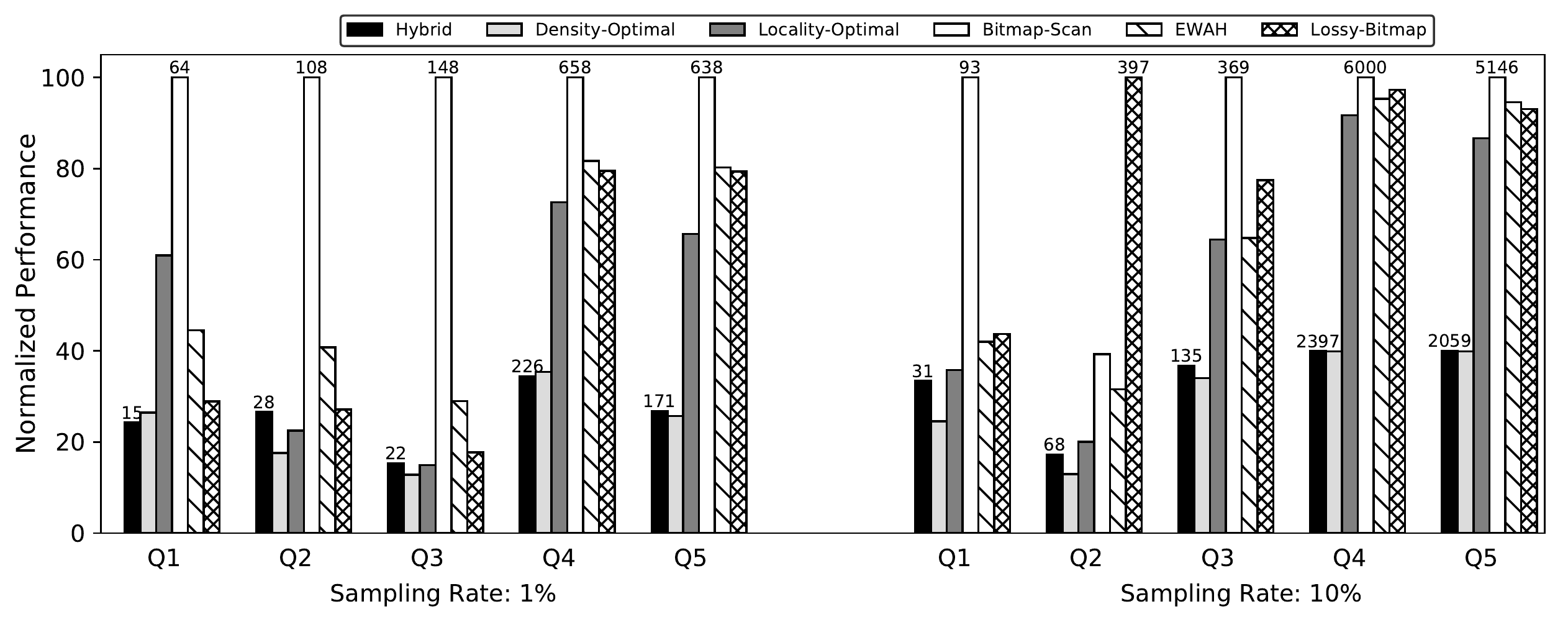}
  \vspace{-10pt}
  \caption{Query runtimes for airline workload on a HDD.}
  \label{fig:airline}
  \vspace{-10pt}
\end{figure*}

\begin{figure*}[tb]
  \centering
  \includegraphics[scale=0.5]{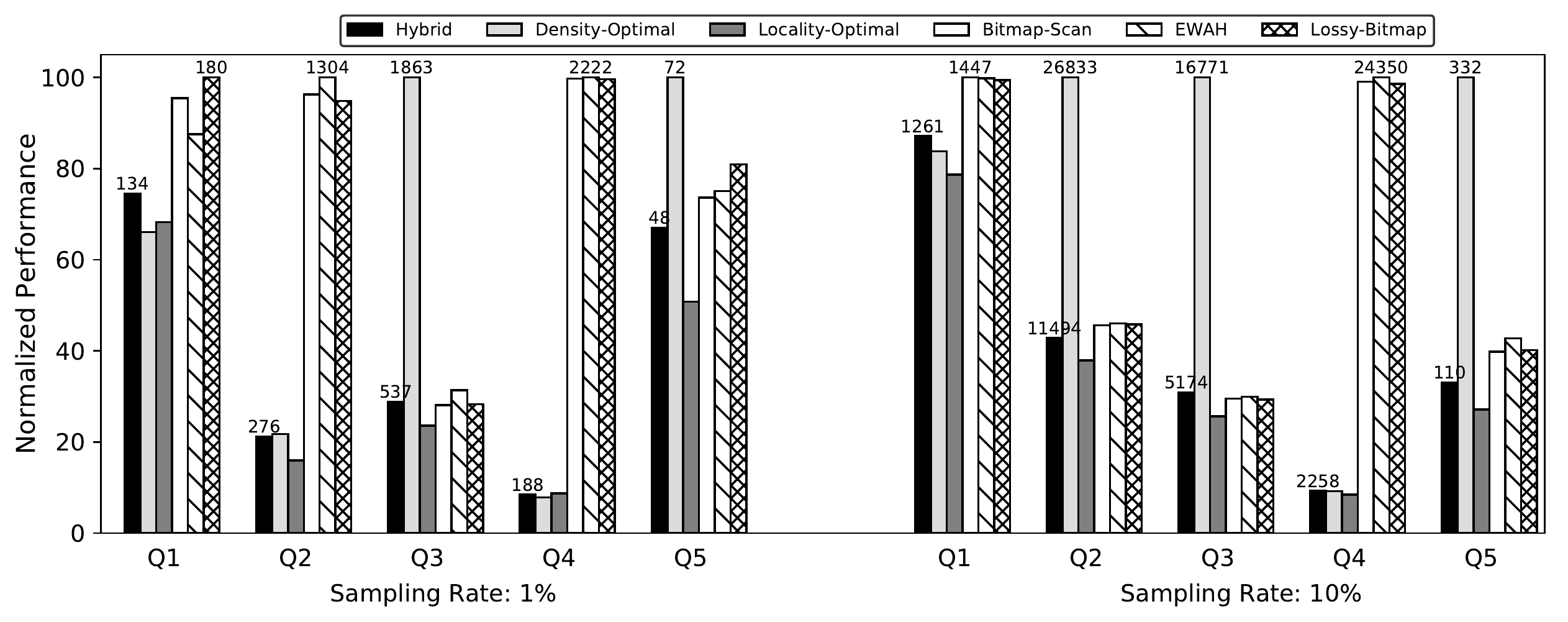}
  \vspace{-10pt}
  \caption{Query runtimes for taxi workload on a HDD.}
  \label{fig:taxitime}
  \vspace{-10pt}
\end{figure*}

\subsection{Query Execution Time}
\label{subsec:exc_time}
\fbox{\begin{minipage}{25em} \small
    \textit{Summary:} In the synthetic datasets on a HDD, our \hybridactual \anyk sampling
    algorithm was on average {\bf 13$\times$} faster than the baselines. For
    the real datasets, \hybridactual performed at least as well
    as the baselines for every query, and on average was {\bf 4$\times$} and
    {\bf 9$\times$} faster for
    queries on HDDs and SSDs respectively. 
\end{minipage}}

\begin{figure*}[tb]
  \centering
  \vspace{-20pt}
  \includegraphics[scale=0.5]{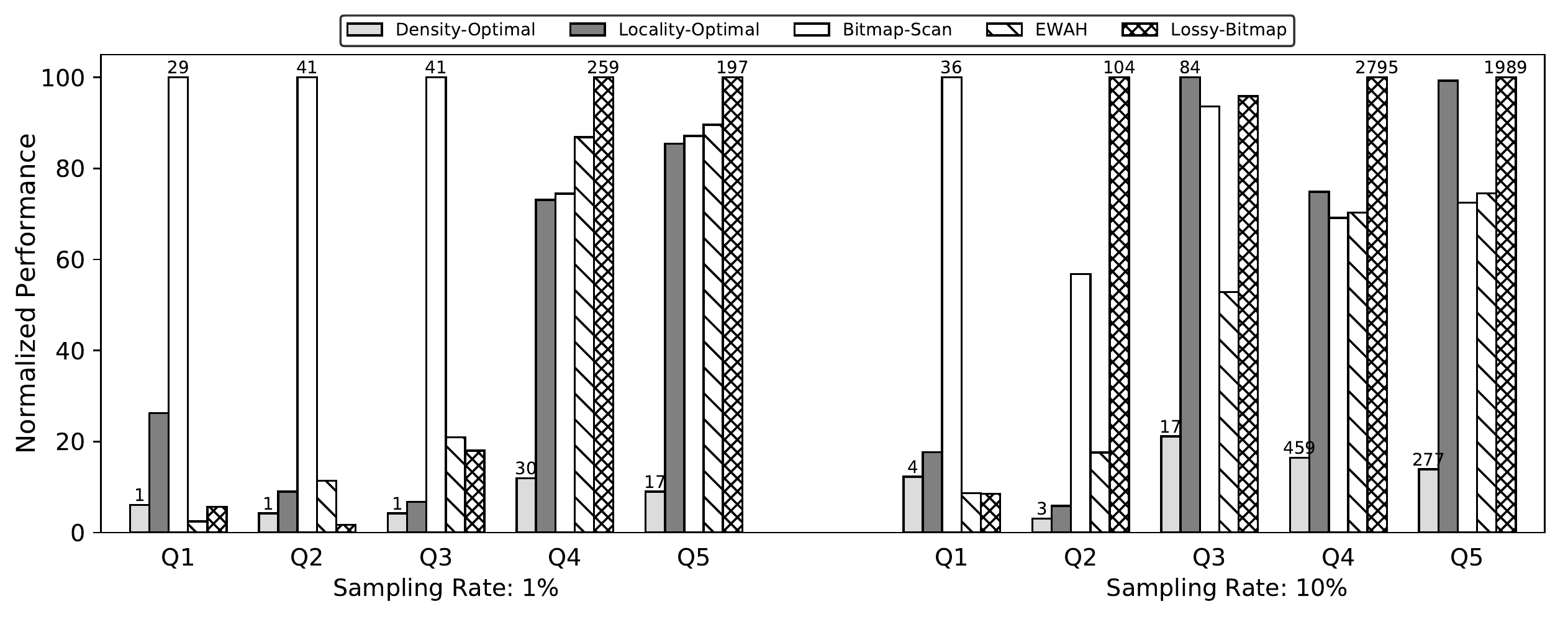}
  \vspace{-10pt}
  \caption{Query runtimes for airline workload on a SSD.}
  \vspace{-5pt}
  \label{fig:flightssd}
\end{figure*}

\techreport{
\begin{figure*}[tb]
  \centering
  \vspace{-5pt}
  \includegraphics[scale=0.5]{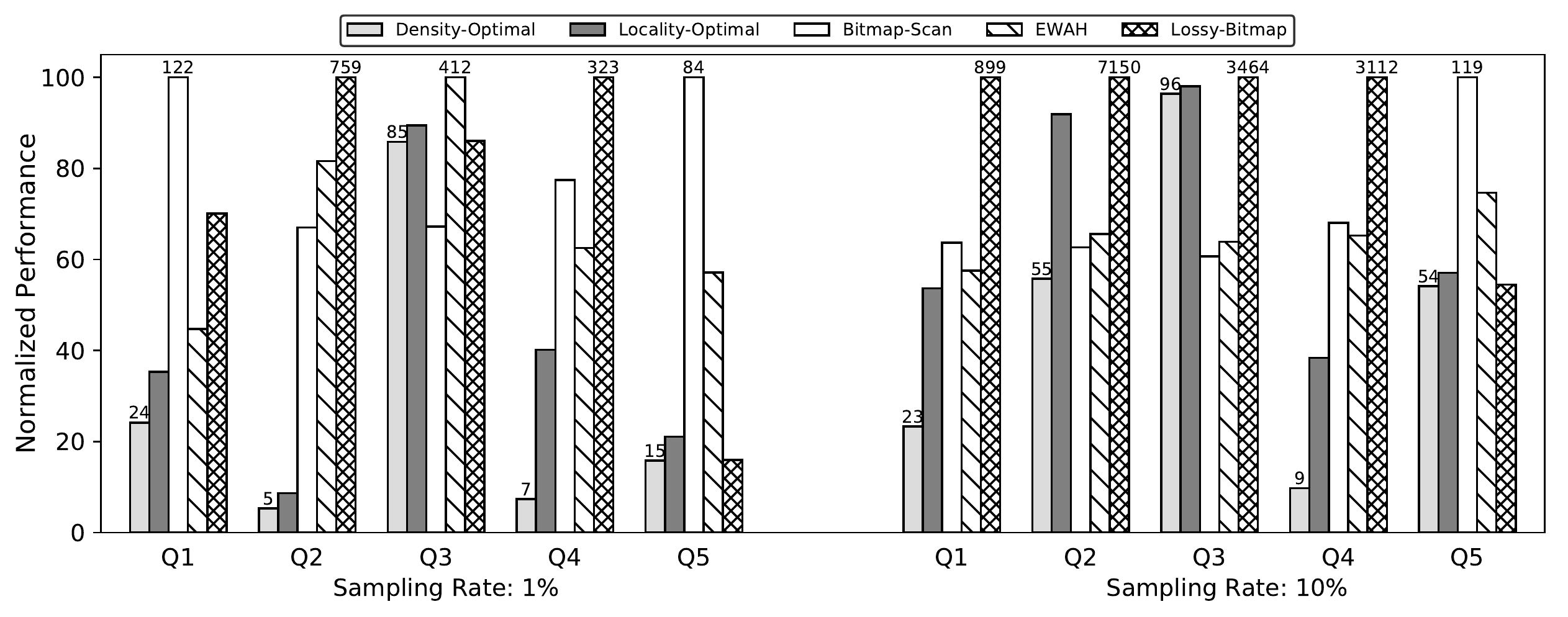}
  \vspace{-10pt}
  \caption{Query runtimes for taxi workload on a SSD.}
  \vspace{-5pt}
  \label{fig:taxissd}
\end{figure*}
}




\stitle{Synthetic Experiments on a HDD} 
Figure~\ref{fig:syndata} presents the runtimes for \hybridactual, \thresh, \prong, and
the four baselines for varying sampling rates. 
(We will evaluate \optimal later on.)
Sampling rate is defined to be the ratio of $k$ divided by
the number of valid records.
Since the queries can have a wide variety in the number of valid records,
we decided to plot the impact on varying sampling rate rather than $k$.
(Results for varying $k$ are similar.)
The bars in the figure above represent the average runtimes for five sampling rates
over 10 synthetic datasets. Note that the figure is in log-scale. 

Regardless of the sampling rate, \thresh, \hybridactual, 
and \prong significantly
outperformed \bmscan, \lossy, \ewah, and \scan, with speedups of an order of magnitude.
For example, for a sampling rate of 1\%, \thresh, \prong, and \hybridactual took 
74ms, 45ms, and 58ms on average respectively,
while \bmscan, \lossy, \ewah, and \scan took
647ms, 624ms, 662ms  and 630ms on average respectively.
Thus, our \anyk algorithms are more effective
at identifying the right sequence of blocks that
contain valid records than the baselines
which do not optimize for \anyk---the
baselines are subject to the vicissitudes of random chance: if
there are large number of valid records early on, then they will do 
well, and not otherwise.
This is despite the fact that \bmscan and \ewah
store more fine-grained information than our algorithms,
and are therefore able to effectively skip over blocks
without valid records.

There was no consistent winner between \thresh and \prong across sampling rates
and queries, but \hybridactual always selected the faster algorithm from the two 
and thus had an average speedup of 13$\times$ over
the baselines.
Despite that, \hybridactual's performance on lower sampling rates (0.1\%, 1\%) is 
a bit worse than \thresh and \prong, since it has to
run both algorithms and pick the better one:
but this difference in performance is small---around 10ms.
From 5\% onwards, \hybridactual's performance is clearly better than \thresh and \prong,
since the increase in computation time is dwarfed by the improvement in 
I/O time.


\stitle{Real Data Experiments on a HDD}
Figures \ref{fig:airline} and \ref{fig:taxitime} show the runtimes of our
algorithms over 5 diverse queries for the airline and taxi workloads respectively. 
For each query and sampling rate, 
we normalized the runtime of each algorithm by the largest
runtime across all algorithms, while also reporting  
actual runtime (in ms) taken by \hybridactual 
and the maximum runtime. We omitted \scan  
since \scan was found to be have the worst runtime in the previous experiment,
and similarly performs poorly here.
For the real
workloads, we noticed that the runtimes of the queries were much more varied,
so we report the average runtime for each query separately. 

For the airline workload, we noticed
that our \anyk algorithms consistently outperformed the
bitmap-based baselines: \thresh
had a speedup of up to 8
compared to \bmscan and \ewah, while \prong
had a speedup of up to 7$\times$. 
Across all queries, when sampling rate equals 1\%, 
\thresh and \prong were on average 3$\times$ and 5$\times$ faster than \bmscan and \ewah,
despite having a much smaller memory footprint (Section~\ref{sec:memory}).
For example for Q3, which had two predicates on month and origin airport, 
the block with the highest density contained 1\% samples already. 
Moreover, since the airline dataset is naturally sorted on time attributes (e.g., year, month), 
the valid tuples were more likely to clustered in a few number of blocks. 
Therefore, compared with \prong,  \thresh fetched up to 10\% less blocks, 
resulting in less query execution time than \prong in all of cases.
For the small additional cost of estimating the
sequence of blocks for both \prong and \thresh, \hybridactual ends up
always selecting the
faster algorithm in both this and the taxi workload,
with an average speedup of 4$\times$. 
For example, for Q4 with 1\% sampling rate \hybridactual's time
is closer to \thresh, and half of that of \prong. 


We noticed a different (and somewhat surprising) trend for the taxi workload. 
Here, \hybridactual continued to do well, and much better than the worst algorithm
on every setting, with an average speedup of 4$\times$
compared to the baselines.
Similarly, \prong performed similar or better than the baselines for every experiment.
However, on multiple occasions, we found that \thresh was slower than the baselines,
and was the worst algorithm, e.g., in Q3 and Q5.
Upon closer examination, we found that \thresh did in fact retrieve the fewest number of
blocks for every query. However, the taxi dataset was much larger than the
airline dataset, so the blocks were more spread out, and the time to seek from
block to block went up significantly. As a result, we found the
locality-favoring \prong to perform better on a HDD where seeks were expensive. 
To further exacerbate the issue, we found that the taxi workload also had a much more uniform
distribution of tuples; the tuples that satisfied query predicates (which were
not based on taxi type)
were spread fairly uniformly across the dataset.
In some sense, this made the dataset ``adversarial'' for density-based 
schemes. 
In other words, it is hard to conclude either \thresh or \prong is 
better than another, given their performance depends on the distribution of valid tuples 
of a given ad-hoc query---and it is therefore safer to use \hybridactual to
pick between the two.



\stitle{Real Data Experiments on a SSD}
We also ran the same workload on SSD; SSDs have 
random I/O performance that is comparable to sequential I/O performance. 
The results are depicted in 
\techreport{Figures~\ref{fig:flightssd} and \ref{fig:taxissd}}\papertext{Figure~\ref{fig:flightssd}
for the airline workload; the taxi workload results are similar
and can be found in~\cite{needletail-tr}}.
We omit \hybridactual, since \hybridactual always selects
\thresh over \prong due to the fact that \thresh fetches
the smallest number of blocks.
Overall, the performance of \thresh is much faster
than the bitmap-based baselines, with average speedups of 
14$\times$ and 6$\times$ in the airline and taxi
workload respectively. 
There were two exceptions: Q1 (10\%) in airline and Q3 in taxi,
where the total number of blocks fetched by \thresh, \bmscan, \lossy,
and \ewah were similar.
In this uncommon situation, even though \thresh has the lowest
I/O time, the CPU cost of checking for valid records in each block was slightly
higher, thus its runtime was a little higher than \bmscan and \ewah. 

\prev{To further explore this issue, 
we reran the experiments on a SSD, which had
random I/O performance that  was comparable to sequential I/O performance. 
 Figure~\ref{fig:taxissd} shows the results. On the SSD, since \thresh fetched
the fewest blocks, it was always faster than \prong. Compared to 
the baselines, \thresh was faster in most of the cases. One exception we
observed was 
in Q3, where the total number of fetched blocks was large, and the number of blocks fetched
by \thresh, \bmscan, \lossy, and \ewah were all similar. 
In this uncommon situation, even though \thresh had the 
lowest I/O time, the CPU cost of checking for valid records in each block was slightly
higher, thus its runtime was a little higher than \bmscan and \ewah for 10\%
sampling rate.}

\subsection{Memory Consumption}
\label{sec:memory}
\fbox{\begin{minipage}{25em} \small
        \textit{Summary:} \densitymaps consumed on average
        {\bf 48$\times$} less memory than the regular bitmaps and
       {\bf 23$\times$} less memory than EWAH-compressed bitmaps. 
\end{minipage}}

\begin{table*}[ht]
\vspace{0pt}
 \begin{center}
 \scalebox{0.8}{    
  \begin{tabular}{|l|c|c|c|c|c|c|c|c|} \hline
	\textbf{Dataset}  & \textbf{Disk Usage}& \textbf{\# Tuples} & \textbf{Cardinality} &\textbf{Bitmap} &\textbf{EWAH} &\textbf{LossyBitmap} &\textbf{DensityMap} \\ \hline
 	Synthetic                &7.5 GB                          & 100M                               &         16          & 190.73MB                 &182.74MB               &  0.06MB                                 &	3.73MB            \\	\hline
	Taxi                 &21 GB                            & 253M                               &         64        &  1936.99MB             &663.63MB              &  0.65MB                                    &41.63MB              \\	\hline
	Airline                      &11GB                             & 123M                               &        805        & 11852.33MB            &744.05MB         & 3.98MB                          &254.72MB           \\	\hline 
     \end{tabular}
 }
 \vspace{-10pt}\caption{Memory consumption of index structures. }
  \label{tab:memoryconsmp}
  \end{center}
\vspace{-15pt}
\end{table*}

\noindent Table~\ref{tab:memoryconsmp} reports the amount of memory used by
\densitymaps compared to the other three bitmap baselines. 
We observed that \densitymaps were 
very light\-weight and
consumed around 51$\times$, 47$\times$, and 47$\times$
less memory than uncompressed bitmaps respectively in the three datasets.
Even with EWAH-compression, we observed an almost 49$\times$ reduction in size
for the taxi dataset for \densitymaps relative to \ewah. 
In the airline dataset, since the selectivity of each attribute value is low, 
\ewah compressed the bitmaps much better than in the other two datasets. Still,
\ewah consumed 3$\times$ more memory than \densitymap. 
Lastly, since \lossy requires only one bit per block while \densitymap is represented as a 
64-bits {\tt double} per block respectively, 
\lossy unsurprisingly consumed less memory than \densitymap. 
However, as we showed in Section \ref{subsec:exc_time}, the smaller memory
consumption incurred a large cost in query latency due to the large number of
false positives (e.g., Q3 with sampling rate 10\% in Figure\ref{fig:airline}); 
especially when the number of predicates is large and exhibit complex correlations.
In comparison, the \densitymap-based \anyk algorithms were orders of
magnitude faster than the baselines, while still maintaining a modest memory footprint
(\mbox{$\sim 0.1\%$} of original dataset).

%
%

\subsection{{\large \optimal} Performance}
\label{subsec:opt-eval}
\fbox{\begin{minipage}{25em} \small
    \textit{Summary:} \optimal had up to {\bf 3.9$\times$} faster I/O time than
    \hybridactual and the best I/O performance among all the algorithms described
    above. However, its large computational cost made it impractical for real
    datasets.
\end{minipage}}

\begin{figure}[htb]
\centering
\vspace{-5pt}
\includegraphics[scale=0.5]{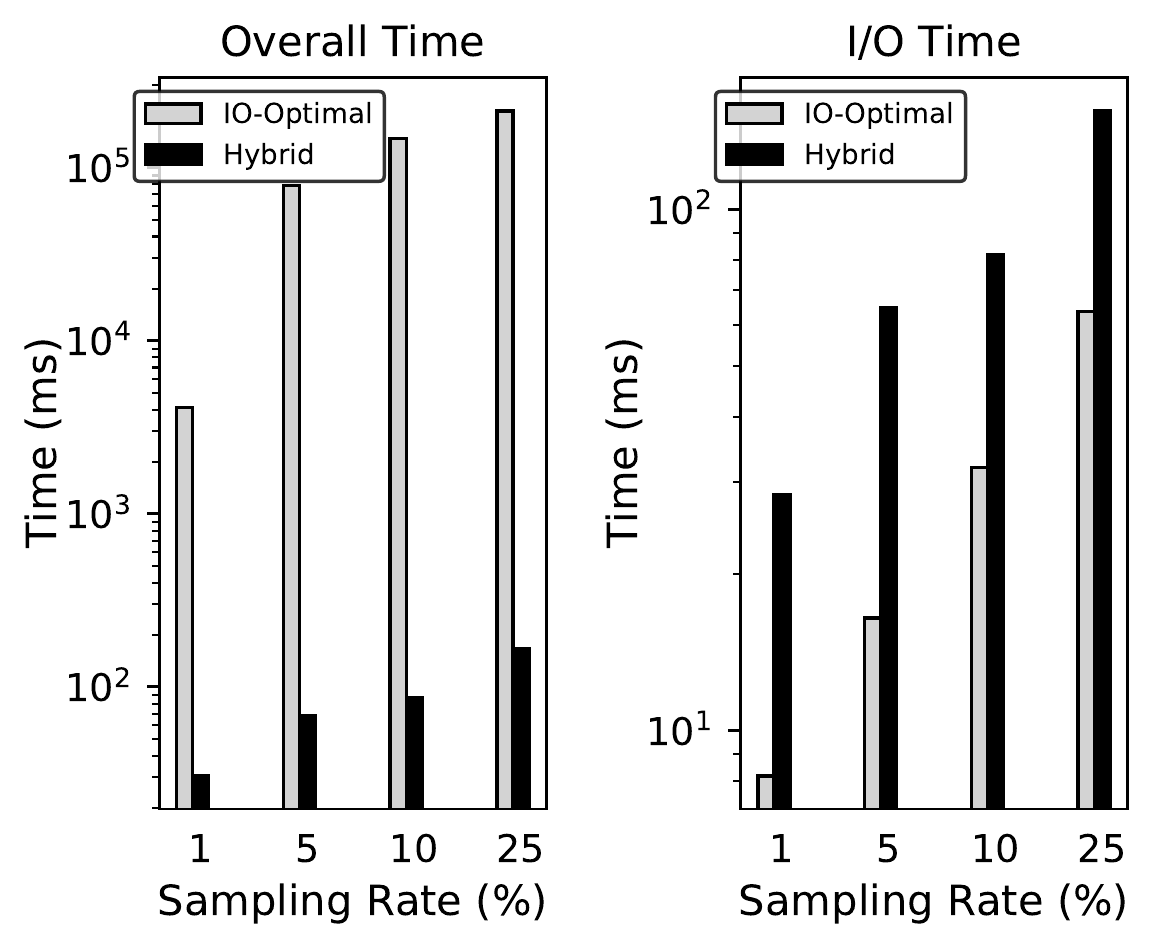}
\vspace{-10pt}
\caption{Overall and I/O time for \optimal and \hybridactual.}
\label{fig:optimal}
\vspace{-10pt}
\end{figure}

\noindent For the evaluation of \optimal, we used a smaller synthetic dataset of 1
million records and a block size of 4KB, and conducted the evaluation on a HDD. 
We compared its overall end-to-end
runtime, CPU time, and I/O time with every other algorithm, and found that it
consistently had the best I/O times. However, we found that computational cost 
of dynamic programming in \optimal outweighed any benefits from the
shorter I/O time. Consequently, we found \optimal to be impractical for 
larger datasets.
Figure~\ref{fig:optimal}
shows both the overall times and I/O times for \optimal and \hybridactual for varying
sampling rates. 

%

\later{
\textbf{Discussion}: 
One modification we can make to \optimal to make it faster
is to further discretize the $k$ values in the dynamic program. Currently, the
dynamic program requires a search over all possible $k$ values. However, if we
only allowed $k$ to go up increments of 100, the search space would be
drastically reduced. This could cause \optimal to speed up by 100$\times$ at
the cost of very little additional I/O time.
}

\subsection{Time vs Error Analysis}\label{subsec:time_err}
\fbox{\begin{minipage}{25em} \small
    \textit{Summary:} Compared to random sampling using bitmap indexes, our \hybrid 
    sampling schemes that mix samples from \anyk sampling algorithms with a
    small percentage of random cluster samples attained the same error rate in much
    less time. 
\end{minipage}}

\begin{figure*}[ht]
  \centering
  \vspace{-10pt}
 
  \subfloat[NYC-Taxi - Ratio Estimator]{
  \includegraphics[scale=0.3]{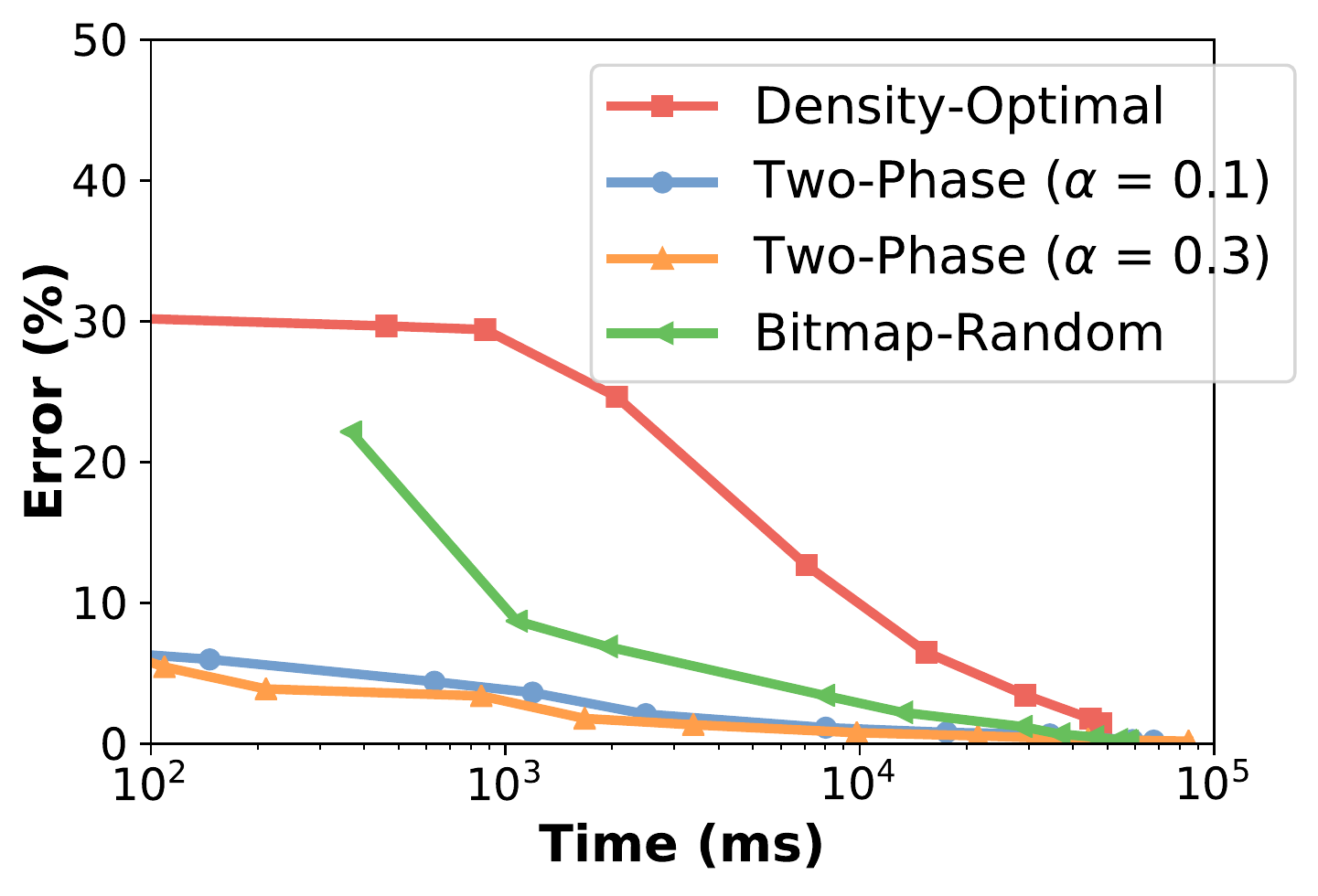}
  \label{fig:taxi-ratio}
  }
  \subfloat[Airline - Ratio Estimator]{
  \includegraphics[scale=0.3]{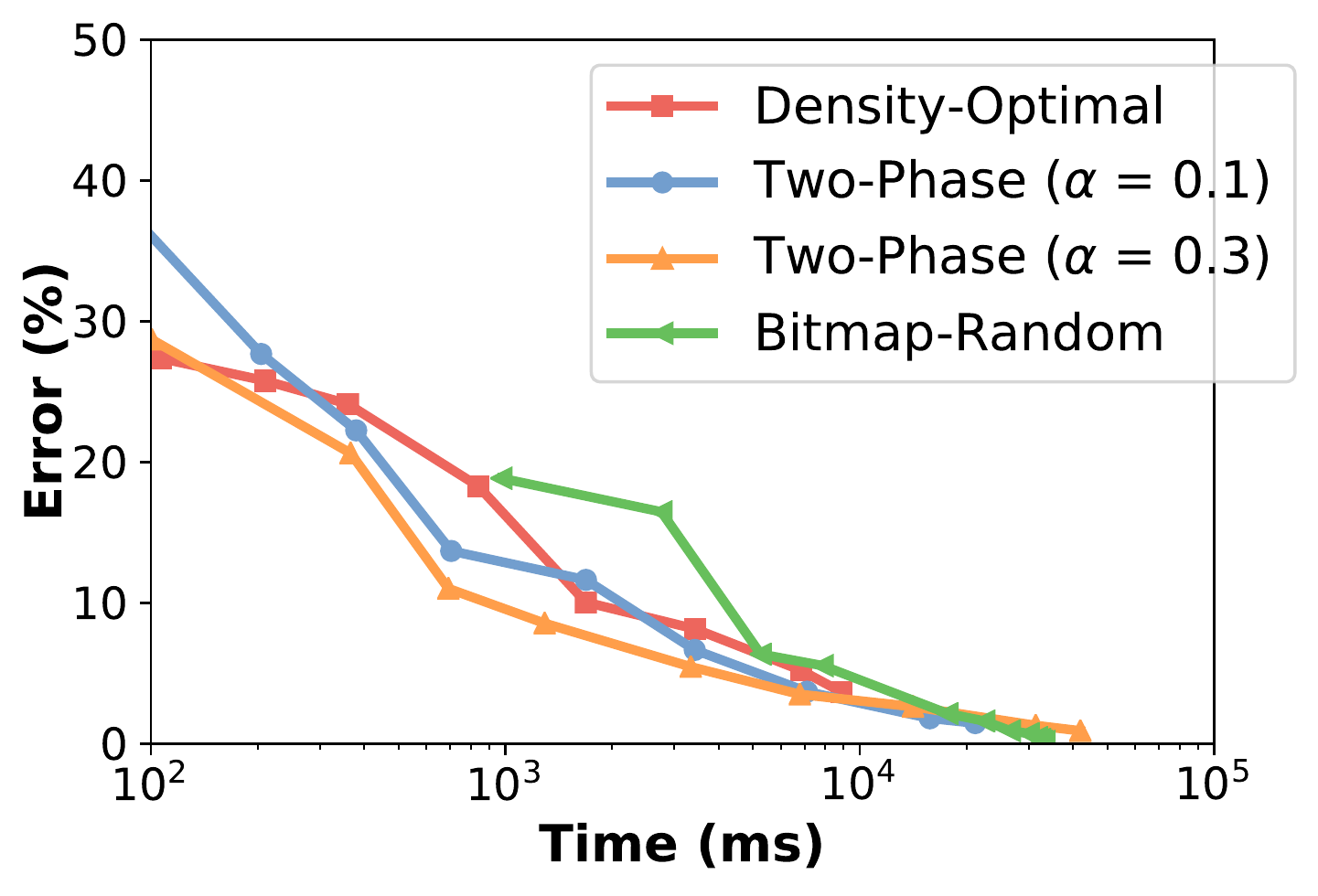}
  \label{fig:airline-ratio}
  } 
  \subfloat[NYC-Taxi - HT Estimator]{
  \includegraphics[scale=0.3]{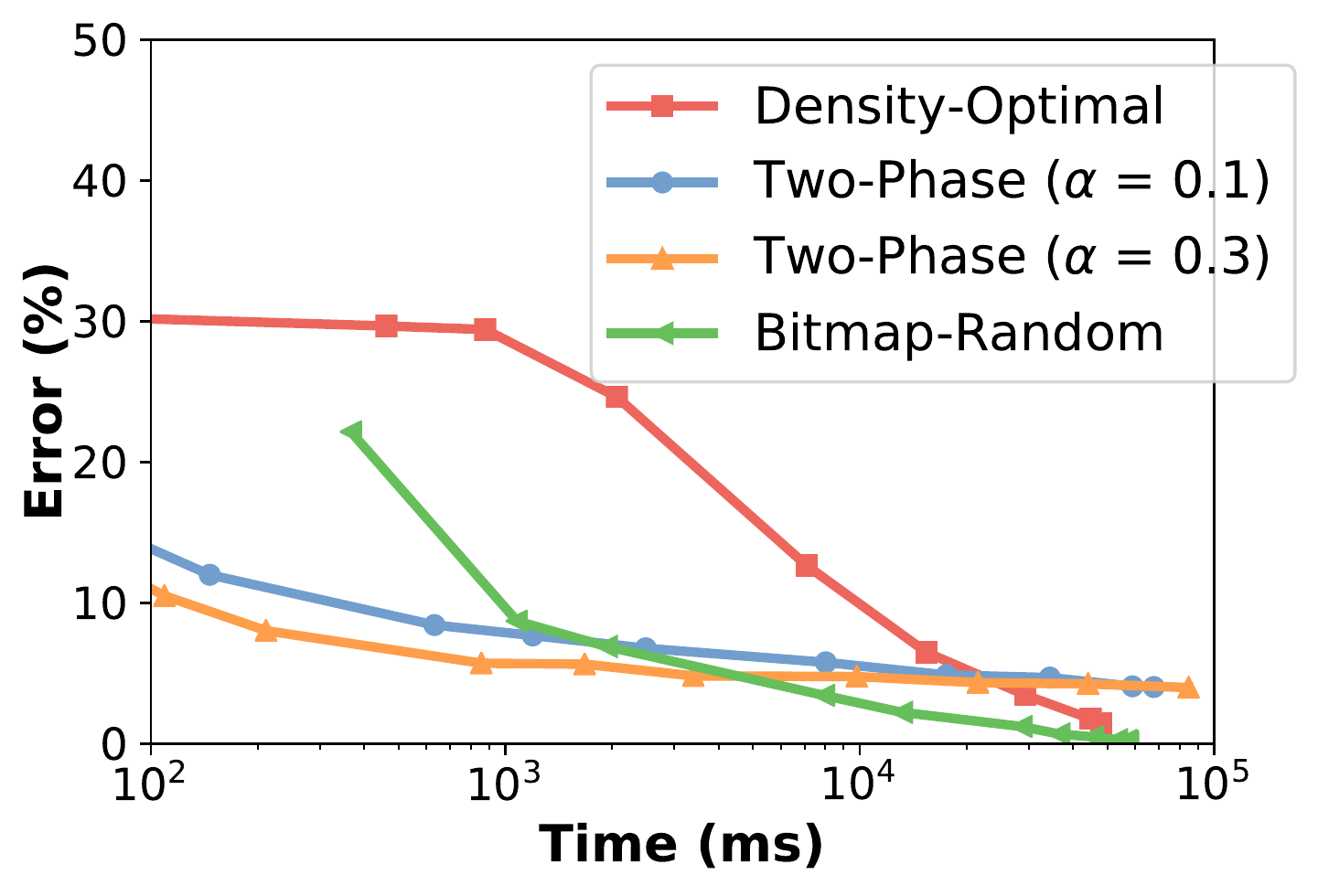}
  \label{fig:taxi-ht}
  }
  \subfloat[Airline - HT Estimator]{
  \includegraphics[scale=0.3]{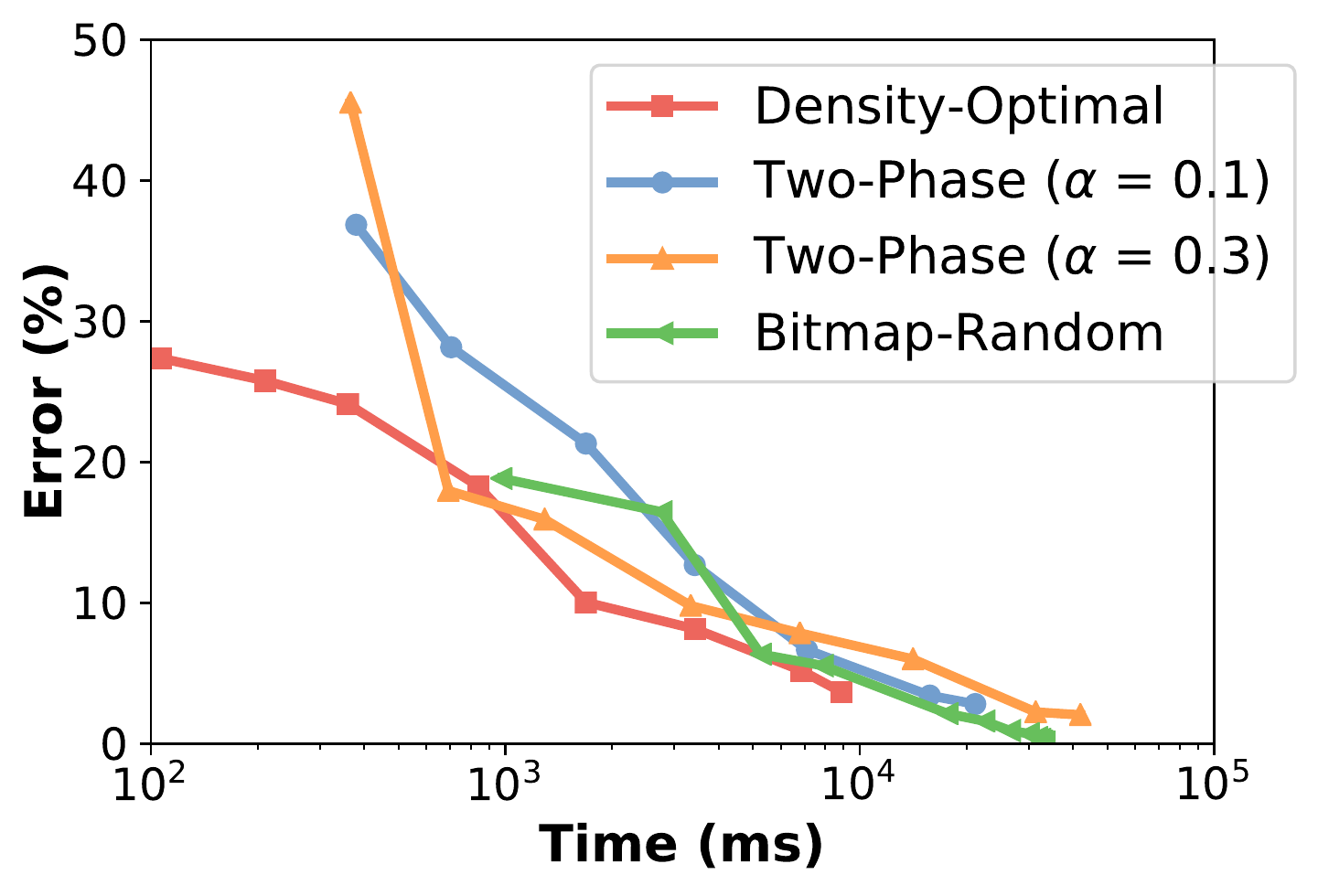}
  \label{fig:airline-ht}
  }
 \vspace{-10pt}
  \caption{Time vs empirical error. 
}
  \label{fig:errorvstime}
\vspace{-8pt}

\end{figure*}


\noindent Using the \hybrid sampling techniques in Section~\ref{sec:est}, we can obtain estimates
of aggregate values on data;  here we experiment with $\alpha= 0\%, 10\%, 30\%$
random samples, and use the \thresh algorithm,
since it
ended up performing the most consistently well across queries and workloads,
for SSDs and HDDs.
We compared
these results with pure random sampling (\bmrand) using bitmaps on a HDD.
We used the same set of queries as in Section~\ref{subsec:exc_time}. For each
query, we varied the sampling rate and measured the runtime and the empirical
error of the estimated aggregate with respect to the true average value.
Figure~\ref{fig:errorvstime} depicts the average results for both the
Horvitz-Thompson estimator and the ratio estimator.
In log
scale.
We'll start with the taxi dataset and the ratio estimator. 
Figure~\ref{fig:taxi-ratio} shows that if all the sampling
schemes are allowed to run for 500ms (commonly regarded as the threshold for interactivity), 
\thresh, \hybrid sampling with $\alpha=0.1$,
\hybrid sampling with $\alpha=0.3$, and \bmrand have average empirical error
rates of 29.64\%, 4.83\%, 3.66\% and 19.64\%, respectively; the corresponding 
number of the samples retrieved are 11102, 7977, 5684, 35 respectively.
Thus, the \hybrid sampling schemes are able to effectively {\em correct the bias}
in \thresh, while still retrieving a comparable amount of samples.
Furthermore, note that \bmrand suffers from the same problem as \bmscan in
large memory
consumption. In contrast,
even though \thresh was not the fastest algorithm in
the taxi workload, our \hybrid sampling algorithms  cluster
sample at the block level and only need access to the much more 
compressed \densitymaps.

The behavior on the airline workload is somewhat different:
here we find that \thresh performs better than the \hybrid sampling scheme
with the ratio estimator for the initial period until about 100ms,
after which the \hybrid sampling schemes perform better than \thresh
and \bmscan. 
We found this behavior repeated across other queries and trials:
\thresh sometimes ends up having very low error (like in Figure~\ref{fig:airline-ratio}),
and sometimes fairly high error (like in Figure~\ref{fig:taxi-ratio}),
but the \hybrid sampling schemes consistently achieve
low error relative to \thresh.
This is because \thresh's accuracy is highly dependent on the
correlation between the data layout and the attribute of interest,
and can sometimes lead to highly biased results. 
At the same time, the \hybrid sampling schemes return
much more samples
 and much more accurate estimates 
than \bmrand, effectively supporting
browsing and sampling at the same time.


Between the Horvitz-Thompson estimator and the ratio estimator, the ratio estimator often
had higher accuracies. 
As explained in Section~\ref{sec:est}\papertext{ and our technical report~\cite{needletail-tr}}, the ratio
estimator works quite well in situations where aggregation estimate is not
correlated with the block densities.
We found this to be the case for both the
airline and taxi workloads, so the ratio estimator helped for both these
workloads.

\techreport{\subsection{Effect of Parameters}
\label{subsec:params}

To explore the properties of our \anyk algorithms, we varied various parameters and
noted their effect on overall runtimes for synthetic workloads. 
Varied parameters included:
\begin{inparaenum}[\itshape (i)\upshape]
\item data size,
\item number of predicates,
\item density,
\item block size, and
\item granularity.
\end{inparaenum}
}

\techreport{
\noindent \textbf{Data Size:}
We varied the synthetic dataset size from 1 million to 1 billion, but we found
that the overall runtimes of our \anyk algorithms remained relatively the
same. Our algorithms return only a fixed $k$ number of samples and
explicitly avoid reading the entire dataset, so it makes sense that the
runtimes stay consistent even when the data size increases.

\noindent \textbf{Number of Predicates}:
As we increased the number of predicates in a query, we saw that overall
runtimes increase as well. Since our predicates were combined using ANDs, an
increase in the number of predicates meant a decrease in the number
of valid records per block. Therefore, both \thresh and \prong needed to fetch more
blocks to retrieve the same number of samples, and this caused an increase in
the overall runtime.




\noindent \textbf{Density}:
As we increased the overall density of valid records in the dataset, the
runtimes for our \anyk algorithms got faster. As the overall density increased,
the average density per block also increased, so our \anyk algorithms could
retrieve fewer blocks to achieve the same number of $k$ samples.




\noindent \textbf{Block Size:}
We tried varying the block sizes of our datasets from 4KB, to 256KB, to 1MB, to
2MB. We found that as we decreased the block sizes, the runtimes for \thresh
increased drastically because smaller block sizes meant that more random I/O
was being done. However, we did not see any definite correlation as we
increased the block size. Although larger block sizes do bias the algorithms
toward more locality, they also mean density information is collected at a
coarser granularity. We suspect that this tradeoff prevented us from seeing any
improvements in performance with increased block size.

}

\subsection{Key-Foreign Key Join Performance}\label{subsec:join_exp}


\papertext{
\begin{table*}[th!]
  \begin{center}
    \scalebox{0.8}{%
      \begin{tabular}{|c|c|c|c|c|c|}
        \hline
        \textbf{Sampling Rate} & \textbf{\shared} & \textbf{\bitcomb} &
        \textbf{\denscomb ($\Psi$=5)} &
        \textbf{\denscomb ($\Psi$=10)} &
        \textbf{\denscomb ($\Psi$=50)} \\ \hline
        0.1\% & 2465.67 & 215.678 & 67.6398 \; (3.19$\times$) & \textbf{63.7114} \; (3.38$\times$) & 73.7204 \; (2.93$\times$)\\ \hline
        0.5\% & 2632.32 & 1049.92 & 307.534 \; (3.41$\times$) & \textbf{300.023} \; (3.50$\times$) & 301.193 \; (3.49$\times$)\\ \hline
        1.0\% & 2761.74 & 1424.12 & 616.090 \; (2.31$\times$) & 593.258 \; (2.40$\times$) & \textbf{591.201} \; (2.41$\times$)\\ \hline
      \end{tabular}
    }
        \vspace{-10pt}
    \caption{%
      Query runtimes (in ms) and speedups compared to \bitcomb for join operations on SDD.
    }
    \vspace{-15pt}
    \label{tab:join_exp}
  \end{center}
\end{table*}
}

\techreport{
\begin{table*}[th!]
  \begin{center}
    \scalebox{0.8}{%
      \begin{tabular}{|c|c|c|c|c|c|}
        \hline
        \textbf{Sampling Rate} & \textbf{\shared} & \textbf{\bitcomb} &
        \textbf{\denscomb ($\Psi$=5)} &
        \textbf{\denscomb ($\Psi$=10)} &
        \textbf{\denscomb ($\Psi$=50)} \\ \hline
        0.1\% & 2465.67 & 215.678 & 67.6398 & \textbf{63.7114} & 73.7204 \\ \hline
        0.5\% & 2632.32 & 1049.92 & 307.534 & \textbf{300.023} & 301.193 \\ \hline
        1.0\% & 2761.74 & 1424.12 & 616.090 & 593.258 & \textbf{591.201} \\ \hline
      \end{tabular}
    }
        \vspace{-10pt}
    \caption{%
      Query runtimes (in ms) on SSD for join operations on
      foreign key tables with 10 million rows.
    }
    \vspace{-15pt}
    \label{tab:join_exp}
  \end{center}
\end{table*}

\begin{table*}[th!]
  \begin{center}
    \scalebox{0.8}{%
      \begin{tabular}{|c|c|c|c|c|c|}
        \hline
        \textbf{Sampling Rate} & \textbf{\shared} & \textbf{\bitcomb} &
        \textbf{\denscomb ($\Psi$=5)} &
        \textbf{\denscomb ($\Psi$=10)} &
        \textbf{\denscomb ($\Psi$=50)} \\ \hline
        0.1\% & 1266.06 & 110.63 & 33.59 & \textbf{32.38} & 68.16 \\ \hline
        0.5\% & 1359.80 & 548.88 & 154.34 & 159.45 & \textbf{153.65} \\ \hline
        1.0\% & 1418.73 & 737.64 & 308.00 & \textbf{299.78} & 307.44 \\ \hline
      \end{tabular}
    }
        \vspace{-10pt}
    \caption{%
      Query runtimes (in ms) on SSD for join operations on
      foreign key tables with 5 million rows.
    }
    \vspace{-15pt}
    \label{tab:join_exp_n5e6}
  \end{center}
\end{table*}

\begin{table*}[th!]
  \begin{center}
    \scalebox{0.8}{%
      \begin{tabular}{|c|c|c|c|c|c|}
        \hline
        \textbf{Sampling Rate} & \textbf{\shared} & \textbf{\bitcomb} &
        \textbf{\denscomb ($\Psi$=5)} &
        \textbf{\denscomb ($\Psi$=10)} &
        \textbf{\denscomb ($\Psi$=50)} \\ \hline
        0.1\% & 12415.26 & 1114.80 & 417.18 & 347.93 & \textbf{314.27} \\ \hline
        0.5\% & 13019.96 & 5193.80 & 2068.75 & 1708.97 & \textbf{1476.64} \\ \hline
        1.0\% & 13651.02 & 7143.95 & 4167.87 & 3478.81 & \textbf{2971.74} \\ \hline
      \end{tabular}
    }
        \vspace{-10pt}
    \caption{%
      Query runtimes (in ms) on SSD for join operations on
      foreign key tables with 50 million rows.
    }
    \vspace{-15pt}
    \label{tab:join_exp_n5e7}
  \end{center}
\end{table*}
}

\techreport{
\begin{table*}[th!]
  \begin{center}
    \scalebox{0.8}{%
      \begin{tabular}{|c|c|c|c|c|c|}
        \hline
        \textbf{Sampling Rate} & \textbf{\shared} & \textbf{\bitcomb} &
        \textbf{\denscomb ($\Psi$=5)} &
        \textbf{\denscomb ($\Psi$=10)} &
        \textbf{\denscomb ($\Psi$=50)} \\ \hline
        0.1\% & 1930.30 & 134.84 & 47.62 & \textbf{46.94} & 95.57 \\ \hline
        0.5\% & 2069.74 & 628.99 & \textbf{223.42} & 228.43 & 225.59 \\ \hline
        1.0\% & 2197.38 & 904.28 & 453.56 & 452.34 & \textbf{447.00} \\ \hline
      \end{tabular}
    }
        \vspace{-10pt}
    \caption{%
      Query runtimes (in ms) on SSD for join operations on a foreign key table
      with 5 attributes.
    }
    \vspace{-15pt}
    \label{tab:join_exp_f5}
  \end{center}
\end{table*}

\begin{table*}[th!]
  \begin{center}
    \scalebox{0.8}{%
      \begin{tabular}{|c|c|c|c|c|c|}
        \hline
        \textbf{Sampling Rate} & \textbf{\shared} & \textbf{\bitcomb} &
        \textbf{\denscomb ($\Psi$=5)} &
        \textbf{\denscomb ($\Psi$=10)} &
        \textbf{\denscomb ($\Psi$=50)} \\ \hline
        0.1\% & 7223.28 & 926.99 & 278.35 & 217.47 & \textbf{165.21} \\ \hline
        0.5\% & 7348.53 & 4460.22 & 1422.41 & 1108.15 & \textbf{886.90} \\ \hline
        1.0\% & 7573.63 & 5758.62 & 2859.31 & 2220.21 & \textbf{1770.48} \\ \hline
      \end{tabular}
    }
        \vspace{-10pt}
    \caption{%
      Query runtimes (in ms) on SSD for join operations on a foreign key table
      with 50 attributes.
    }
    \vspace{-15pt}
    \label{tab:join_exp_f50}
  \end{center}
\end{table*}
}

\techreport{
\begin{table*}[th!]
  \begin{center}
    \scalebox{0.8}{%
      \begin{tabular}{|c|c|c|c|c|c|}
        \hline
        \textbf{Sampling Rate} & \textbf{\shared} & \textbf{\bitcomb} &
        \textbf{\denscomb ($\Psi$=5)} &
        \textbf{\denscomb ($\Psi$=10)} &
        \textbf{\denscomb ($\Psi$=50)} \\ \hline
        0.1\% & 2220.28 & 226.89 & \textbf{62.43} & 65.48 & 74.75 \\ \hline
        0.5\% & 2389.59 & 1110.59 & 312.64 & \textbf{303.34} & 304.42 \\ \hline
        1.0\% & 2509.94 & 1478.30 & 625.06 & \textbf{605.06} & 606.28 \\ \hline
      \end{tabular}
    }
        \vspace{-10pt}
    \caption{%
      Query runtimes (in ms) on SSD for join operations on a primary key table
      with 5 attributes.
    }
    \vspace{-15pt}
    \label{tab:join_exp_p5}
  \end{center}
\end{table*}

\begin{table*}[th!]
  \begin{center}
    \scalebox{0.8}{%
      \begin{tabular}{|c|c|c|c|c|c|}
        \hline
        \textbf{Sampling Rate} & \textbf{\shared} & \textbf{\bitcomb} &
        \textbf{\denscomb ($\Psi$=5)} &
        \textbf{\denscomb ($\Psi$=10)} &
        \textbf{\denscomb ($\Psi$=50)} \\ \hline
        0.1\% & 5177.96 & 245.09 & \textbf{70.56} & 75.44 & 85.31 \\ \hline
        0.5\% & 5453.41 & 1172.76 & 361.24 & \textbf{350.30} & 351.10 \\ \hline
        1.0\% & 5692.40 & 1618.83 & 722.81 & \textbf{696.12} & 698.50 \\ \hline
      \end{tabular}
    }
        \vspace{-10pt}
    \caption{%
      Query runtimes (in ms) on SSD for join operations on a primary key table
      with 50 attributes.
    }
    \vspace{-15pt}
    \label{tab:join_exp_p50}
  \end{center}
\end{table*}
}

\techreport{
\begin{table*}[th!]
  \begin{center}
    \scalebox{0.8}{%
      \begin{tabular}{|c|c|c|c|c|c|}
        \hline
        \textbf{Sampling Rate} & \textbf{\shared} & \textbf{\bitcomb} &
        \textbf{\denscomb ($\Psi$=5)} &
        \textbf{\denscomb ($\Psi$=10)} &
        \textbf{\denscomb ($\Psi$=50)} \\ \hline
        0.1\% & 2528.40 & 68.33 & 60.85 & \textbf{58.91} & 68.16 \\ \hline
        0.5\% & 2699.45 & 326.55 & 280.79 & \textbf{278.06} & 279.73 \\ \hline
        1.0\% & 2843.45 & 651.10 & 558.19 & \textbf{547.48} & 559.85 \\ \hline
      \end{tabular}
    }
        \vspace{-10pt}
    \caption{%
      Query runtimes (in ms) on SSD for join operations on
      5 unique join attribute values.
    }
    \vspace{-15pt}
    \label{tab:join_exp_j5}
  \end{center}
\end{table*}

\begin{table*}[th!]
  \begin{center}
    \scalebox{0.8}{%
      \begin{tabular}{|c|c|c|c|c|c|}
        \hline
        \textbf{Sampling Rate} & \textbf{\shared} & \textbf{\bitcomb} &
        \textbf{\denscomb ($\Psi$=5)} &
        \textbf{\denscomb ($\Psi$=10)} &
        \textbf{\denscomb ($\Psi$=50)} \\ \hline
        0.1\% & 2445.25 & \textbf{1250.18} & 2148.35 & 1845.65 & 1603.79 \\ \hline
        0.5\% & 2597.23 & \textbf{1513.81} & 2304.90 & 2008.87 & 1759.71 \\ \hline
        1.0\% & 2757.08 & \textbf{1704.73} & 2421.01 & 2116.32 & 1873.40 \\ \hline
      \end{tabular}
    }
        \vspace{-10pt}
    \caption{%
      Query runtimes (in ms) on SSD for join operations on
      50 unique join attribute values.
    }
    \vspace{-15pt}
    \label{tab:join_exp_j50}
  \end{center}
\end{table*}

\begin{table*}[th!]
  \begin{center}
    \scalebox{0.8}{%
      \begin{tabular}{|c|c|c|c|c|c|}
        \hline
        \textbf{Sampling Rate} & \textbf{\shared} & \textbf{\bitcomb} &
        \textbf{\denscomb ($\Psi$=5)} &
        \textbf{\denscomb ($\Psi$=10)} &
        \textbf{\denscomb ($\Psi$=50)} \\ \hline
        0.1\% & 1527 & 1527 & \textbf{1284} & \textbf{1284} & \textbf{1284} \\ \hline
        0.5\% & 1527 & 1527 & \textbf{1284} & \textbf{1284} & \textbf{1284} \\ \hline
        1.0\% & 1527 & 1527 & \textbf{1284} & \textbf{1284} & \textbf{1284} \\ \hline
      \end{tabular}
    }
        \vspace{-10pt}
    \caption{%
      Number of blocks fetched for join operations on
      50 unique join attribute values.
    }
    \vspace{-15pt}
    \label{tab:join_exp_j50b}
  \end{center}
\end{table*}
}

\techreport{
\begin{table*}[th!]
  \begin{center}
    \scalebox{0.8}{%
      \begin{tabular}{|c|c|c|c|c|c|}
        \hline
        \textbf{Sampling Rate} & \textbf{\shared} & \textbf{\bitcomb} &
        \textbf{\denscomb ($\Psi$=5)} &
        \textbf{\denscomb ($\Psi$=10)} &
        \textbf{\denscomb ($\Psi$=50)} \\ \hline
        0.1\% & 62154.22 & 142.40 & 56.50 & \textbf{54.40} & 78.98 \\ \hline
        0.5\% & 63513.28 & 664.69 & 273.63 & 259.14 & \textbf{252.80} \\ \hline
        1.0\% & 64099.30 & 1308.80 & 542.67 & 522.12 & \textbf{510.73} \\ \hline
      \end{tabular}
    }
        \vspace{-10pt}
    \caption{%
      Query runtimes (in ms) on SSD for join operations on a foreign key
      with Zipf distribution parameter of 1.5.
    }
    \vspace{-15pt}
    \label{tab:join_exp_z1.5}
  \end{center}
\end{table*}

\begin{table*}[th!]
  \begin{center}
    \scalebox{0.8}{%
      \begin{tabular}{|c|c|c|c|c|c|}
        \hline
        \textbf{Sampling Rate} & \textbf{\shared} & \textbf{\bitcomb} &
        \textbf{\denscomb ($\Psi$=5)} &
        \textbf{\denscomb ($\Psi$=10)} &
        \textbf{\denscomb ($\Psi$=50)} \\ \hline
        0.1\% & 1673.12 & \textbf{1077.73} & 1783.76 & 1706.39 & 1647.18 \\ \hline
        0.5\% & 1704.48 & \textbf{1120.75} & 1823.00 & 1742.84 & 1711.86 \\ \hline
        1.0\% & 1742.57 & \textbf{1186.66} & 1841.09 & 1734.92 & 1721.47 \\ \hline
      \end{tabular}
    }
        \vspace{-10pt}
    \caption{%
      Query runtimes (in ms) on SSD for join operations on a foreign key
      with Zipf distribution parameter of 5.
    }
    \vspace{-15pt}
    \label{tab:join_exp_z5}
  \end{center}
\end{table*}

\begin{table*}[th!]
  \begin{center}
    \scalebox{0.8}{%
      \begin{tabular}{|c|c|c|c|c|c|}
        \hline
        \textbf{Sampling Rate} & \textbf{\shared} & \textbf{\bitcomb} &
        \textbf{\denscomb ($\Psi$=5)} &
        \textbf{\denscomb ($\Psi$=10)} &
        \textbf{\denscomb ($\Psi$=50)} \\ \hline
        0.1\% & 1526 & 1526 & \textbf{1525} & \textbf{1525} & \textbf{1525} \\ \hline
        0.5\% & 1527 & 1527 & \textbf{1525} & \textbf{1525} & \textbf{1525} \\ \hline
        1.0\% & 1527 & 1527 & \textbf{1525} & \textbf{1525} & \textbf{1525} \\ \hline
      \end{tabular}
    }
        \vspace{-10pt}
    \caption{%
      Number of blocks fetched for join operations on a foreign key
      with Zipf distribution parameter of 5.
    }
    \vspace{-15pt}
    \label{tab:join_exp_z5b}
  \end{center}
\end{table*}
}

\fbox{\begin{minipage}{25em} \small
    \textit{Summary:} Our iterative join \anyk algorithm has an average
    speedup of {\bf 3$\times$} compared to existing baselines not optimized for \anyk.
\end{minipage}}

We now evaluate the extension of our \anyk algorithms to key-foreign key joins
from Section~\ref{sec:join}.
We compare the performance of our join algorithm with two baselines: (1)
\shared: a single scan of the foreign key table, shared
across different join attribute values,  with no indexes and (2)
\bitcomb: a single scan of the foreign key table, shared across different 
join attribute values, with bitmap indexes to
skip to the next valid record which can serve as a sample.
In both cases, the algorithms terminate as soon as $k$ samples for each join value
are found, and a hash join is used to combine the foreign key record with the
primary key record, with the hash table constructed on the primary key table.
In \bitcomb, bitmaps for different join values were first combined using OR,
then once a join attribute value had reached $k$ samples, its bitmap is subtracted from
the combined bitmap.

For our join \anyk algorithm, we ran the iterative algorithm presented in
Section~\ref{sec:join} and used \thresh as our \anyk algorithm in each
iteration. We varied the number of blocks retrieved per iteration ($\Psi$)
before the updates to the combined densities, and evaluated its impact on the
overall runtime. As with \shared and \bitcomb, we used a hash join with the
hash table constructed on the primary key table.

All experiments were run with a synthetic dataset using a SSD drive.
The synthetic dataset had two tables: one for the primary key and one for the
foreign key.
All attributes for both tables were of \texttt{int} type.
The primary key table's $i$th row had $i$ as the its unique primary key value,
and the foreign key
table's foreign key attribute values were generated using a Zipf distribution.
Note that this means there were some foreign keys which did not match with any
primary key.
In addition, we varied the following parameters:
\begin{inparaenum}[(1)]
\item the number of rows in the foreign key table,
\item the number of attributes in the foreign key table,
\item the number of attributes in the primary key table,
\item number of unique values for the join attribute in the primary key table
(and thereby the number of the rows in the primary key table), and
\item the Zipf distribution parameter.
\end{inparaenum}
\techreport{Each experimental setup was run 5 times and the means of these
  average runtimes are reported in this paper. The standard deviation between
  the runtimes were less than 1\% for the experiments, so they are not
reported.}

Table~\ref{tab:join_exp} shows the overall runtimes for different sampling
rates with the following parameters:
\begin{inparaenum}[(1)]
\item 10 million rows for the foreign key table,
\item 10 foreign key key table attributes,
\item 10 primary key table attributes,
\item 10 unique join attribute values, and
\item 2 as the Zipf distribution parameter.
\end{inparaenum}
The lowest runtimes for each sampling rate are highlighted in bold and the
speedup relative to \bitcomb (the better of the two baselines) are indicated in
parentheses.
As shown, our \denscomb was the fastest algorithm for each
sampling rate, with a 3$\times$ speedup compared to \bitcomb and an order of
magnitude difference with respect to \shared.
This was largely due to the fact that \denscomb retrieved far fewer blocks than
either \bitcomb or \shared. For a sampling rate of 0.05\%, \denscomb ($\Psi$=10)
only retrieved 190 blocks, while \bitcomb retrieved 1259 and \shared retrieved
1527. Since we used a SSD, this always resulted in a lower runtime. Note that
had these experiments been run on a HDD, we would have used \hybridactual as our
\anyk algorithm.

We found that varying $\Psi$ had a rather minimal impact on the overall
runtime for values between a certain range (5 - 50 for this case).
Outside of this range (e.g., $\Psi=1$), $\Psi$ had a larger impact on 
performance, but the overall runtime was still lower than either \bitcomb or
\shared.

\papertext{
We found these observations to hold true for other variations of the
parameters. Only when the number of unique primary key values was large, and
the Zipf distribution parameter large, did we see that \bitcomb and \shared
perform even comparably to \denscomb. More details can be found in our technical
report~\cite{needletail-tr}.
}

\techreport{
  With the experimental setup used for Table~\ref{tab:join_exp}, we varied each
  of the 5 parameters mentioned before one at a time to see their effect on
  overall runtime performance. Other than the varied parameter for each
  experiment, the other parameters were set to be the same as those used in the
  experiment for Table~\ref{tab:join_exp}.

  \stitle{(1) Rows in Foreign Key Table}
  First, we wanted to see whether our iterative join \anyk algorithm could
  scale to different dataset sizes. Since, the size of the primary key table is
  fixed to be the number of unique join attribute values, we first focused on
  varying the number of rows in the foreign key table.
  Tables~\ref{tab:join_exp_n5e6} and~\ref{tab:join_exp_n5e7}
  show the results for 5 and 50 million rows in the foreign key table
  respectively. As we
  can see \denscomb still provided a speedup of 2-3$\times$ over \bitcomb,
  suggesting that our join \anyk algorithm can scale. The reasons for the
  speedup were the same as for Table~\ref{tab:join_exp}; much fewer blocks were
  retrieved by \denscomb than \bitcomb.

  \stitle{(2) Number of Attributes in Foreign Key Table}
  Given the row-oriented layout of our data, a change in the number of
  attributes in the foreign key table meant a change in the number of records
  per block for the foreign table. This parameter allowed us to observe how \denscomb
  would adapt to different numbers of records in the blocks. Furthermore, 
  the number of attributes also affected the size of the dataset, so this
  experiment served as an additional check on how well \denscomb scaled with
  size.
  Tables~\ref{tab:join_exp_f5} and~\ref{tab:join_exp_f50} show the results for
  5 and 50 attributes in the foreign key table respectively. \denscomb still
  remained faster than either \shared or \bitcomb, and we saw that as the
  number of attributes increased, the speedup became more pronounced as well.
  This was due to \denscomb being more selective with the blocks it chose to
  retrieve. When there were fewer records per block, the choice of the blocks
  had a large impact on the number of blocks fetched, making \denscomb more
  suited for this case than either \bitcomb or \shared.

  \stitle{(3) Number of Attributes in Primary Key Table}
  We hypothesized that varying the number of attributes in the primary key
  table would have minimal effect on the overall runtime. The only
  impact this variable should have had was on the time it took to copy the
  record in the primary key table for the output. Tables~\ref{tab:join_exp_p5}
  and~\ref{tab:join_exp_p50} show the results for 5 and 50 attributes in the
  primary key table respectively. As we expected, this parameter did have a
  minimal impact on the overall performance of the algorithms, with runtimes
  extremely similar to Table~\ref{tab:join_exp}. We believe given the
  magnitude of the difference, the runtime discrepancies between
  Table~\ref{tab:join_exp} and~\ref{tab:join_exp_p5} were due to experimental
  noise.

  \stitle{(4) Number of Unique Join Attribute Values}
  We wanted to see how \denscomb would perform with different number of join
  values, so we altered the number the number of unique join attribute values,
  and thereby also increased the number of rows in the primary key table.
  Table~\ref{tab:join_exp_j5} and~\ref{tab:join_exp_j50} show the results for 5
  and 50 unique join attribute values respectively. As expected, \denscomb
  outperformed \bitcomb for 5 unique join values. Interestingly,
  \bitcomb was more performant than \denscomb
  for 50 unique join values.
  Upon closer examination, we found that although \bitcomb was faster than
  \denscomb, \denscomb was still retrieving fewer blocks as shown by
  Table~\ref{tab:join_exp_j50b}. However, compared to the other experiments,
  \denscomb was returning a larger ratio of blocks with respect to \bitcomb.
  Due to the Zipf distribution nature of the foreign key values, records with a
  foreign key value greater than 10 were scarce, and more spread out among the
  blocks. This meant that a greater number of blocks would have to be returned
  to satisfy the users' join \anyk query (regardless of any algorithm used).
  Since we ran the all these experiments with \thresh, we believe that
  \denscomb's lack of awareness caused a greater overall runtime.

  \stitle{(5) Zipf Distribution Parameter}
  The final variable of interest was the Zipf distribution parameter used to
  generate the attribute values for the foreign key.
  Table~\ref{tab:join_exp_z1.5} and~\ref{tab:join_exp_z5} show the results for
  a Zipf distribution parameter of 1.5 and 5 respectively.
  \denscomb outperformed \shared and \bitcomb as usual for a Zipf distribution
  parameter of 1.5, but \bitcomb once again outperformed \denscomb for a Zipf
  distribution parameter of 5. A greater Zipf distribution parameter forced the
  foreign keys to be more heavily concentrated around the lower numbers, thus
  causing the higher numbers to become more scarce. Thus, a similar
  behavior to the experiment with 50 unique join attribute values was
  exhibited. We can see this from Table~\ref{tab:join_exp_z5b}, in which
  \denscomb still retrieved the ``{fewest}'' number of blocks, but it was only
  1 or 2 less than \bitcomb and \shared. When retrieving around the same number
  of blocks, the locality-unaware \denscomb expectedly performed worse than
  \bitcomb.

  \stitle{Overall} These experiments suggest that the iterative join \anyk
  algorithm is most effective when a smaller number of blocks needs to be
  fetched. Luckily, most browsing cases fit into such a category (e.g., a user
  is not likely to want 10 samples for each of the 50 unique join values), so
  \ntail ends up being a good fit for the browsing use case. Nevertheless, we
  are still in the midst of trying to make \denscomb more aware of locality so
  that it can handle any situation.
}

\balance

\section{Related Work}\label{sec:related}

Prior work related to \ntail can be divided into the following categories:

\stitle{Data Skipping} Intelligently identifying and skipping irrelevant blocks
can significantly reduce system I/O time.  For example, OLAP systems use indexes
that track the minimum and maximum value in each block to skip blocks that do
not satisfy queries~\cite{sahuguet2001building, slȩzak2008brighthouse,
lamb2012vertica}. Sun et al.~\cite{sun2014fine} employ a workload-aware version
of this technique that, given common filters in a past workload, partitions data
into multiple blocks and skips irrelevant blocks at runtime.  \ntail also skips
blocks, but \densitymaps require no workload to set up and allow us to quickly
identify which blocks contain the most records, allowing us to develop our \anyk
techniques.

\stitle{Bitmap Indexing} Bitmap indexes~\cite{DBLP:dblp_conf/sigmod/ChanI98} improve response-time for queries with multiple boolean predicates by composing bitmaps to  filter out rows that do not satisfy the query conditions. The key limitation of bitmap indexes is that their size increases significantly as the cardinality of attributes grows. There exist various techniques to reduce the size of bitmaps, including compression~\cite{johnson1999performance,wu2004performance, antoshenkov1996query,sidirourgos2013column}, encoding~\cite{chan1998bitmap, wu2001compressed} and binning~\cite{stockinger2004evaluation, wu2008breaking}.  For the specific problem of \anyk sampling, \densitymaps provide a coarser indexing structure that is smaller, faster and sufficient to identify dense blocks without involving compression or decompression. 

\stitle{Block Level Indexing} This group of indexing techniques,
including \lossy~\cite{wiki:lossybitmap}, SMA~\cite{moerkotte2008small} and
variants of SMA~\cite{lang2016data, boncz2005monetdb, francisco2011netezza}, were
developed to track aggregate attribute information at the block level. These
techniques have been used to aid query processing 
in database systems such as Vertica~\cite{lamb2012vertica},
Netezza~\cite{netezza}, and MonetDB/X100~\cite{boncz2005monetdb}.
By tracking aggregate information, these techniques are able to consume
less memory than finer-grained index structures such as
regular record-level bitmap indexes. 
While our \densitymap also lies in the same family as these
aggregate block-based techniques, \densitymaps are significantly better-suited
for the \anyk problem. The densities in \densitymaps
allow our \anyk algorithms to prioritize blocks which are more likely to have
valid records, thereby significantly reducing the number of
retrieved blocks and overall I/O time.
%
%
%
In
Section \ref{sec:eval},  we experimentally demonstrate that our \anyk algorithms
using \densitymaps outperforms \lossy by up to 5$\times$ and 6$\times$ in HDDs and SSDs respectively.

\stitle{Approximate Query Processing} In the past decade, a number of approximate query processing techniques~\cite{Garofalakis:2001:AQP:645927.672356, gibbons2001distinct, jermaine2008scalable}  and 
systems~\cite{DBLP:conf/eurosys/AgarwalMPMMS13,Acharya:1999:AAQ:304182.304581} have emerged 
that allow users to trade off query accuracy for interactive response times, by employing random sampling. 
These techniques fall into one of two categories:
either they pre-materialize specific samples or sketches of data, tailored to the queries~\cite{DBLP:conf/eurosys/AgarwalMPMMS13,Babcock:2003:DSS:872757.872822,914867,Ioannidis:1999:HAS:645925.671527,acharya1999join},
or perform some form of online sampling~\cite{jermaine2008scalable,online-aggregation,haas1999ripple,wang2015spatial}.
The former category does not apply to exploratory data analysis, since a workload is assumed.
The latter category use techniques that are either similar to \bmrand or \scan in order to achieve
adequate randomization. 
In contrast, \ntail primarily focuses on \anyk sampling, possibly 
extended with random sampling.  This allows \ntail to avoid accessing 
data in random order, avoiding expensive up-front 
randomization or inefficient random access to data at runtime.

\stitle{Random Sampling in Relational Databases} Olken and Rotem \cite{olken1993random} examine data structures, algorithms and their performance for simple random sampling from a variety of relational operators. Various database systems~\cite{ibmrandsampling,msrandsampling} extend SQL with functions that lets users randomly select a subset of rows from the query results. However, since these techniques are based on random sampling, they incur high latency even for retrieving a small (1\%) amount of samples. In \ntail, our \hybrid sampling technique returns much larger samples than random sampling, but in much less time and with comparable accuracy.   

\stitle{Output Rate Maximization}
A related line of work is that of generating join results early, trying to increase
the rate of output of tuples~\cite{tao2005rpj,tok2007rrpj,tok2006progressive,bornea2010adaptive,viglas2003maximizing}. 
In particular, the papers aim to identify the tuples that are most beneficial to preferentially
cache in memory so as to maintain a high output rate,
trading off early join results and end-to-end execution time.
These papers do not formally articulate or optimize the \anyk problem.
Moreover, our approach is instead to preferentially read certain portions of the data
to solve the \anyk problem; thus our approaches are complementary.
\papertext{We provide more details in our technical report~\cite{needletail-tr}.}


\techreport{In particular, RPJ~\cite{tao2005rpj} formulates the output rate based on the data distribution and develops an optimal flush policy when input exceeds memory budget. Instead, RRPJ~\cite{tok2007rrpj} directly observes the output rate and flushes data according to result statistics. Wee et al.,~\cite{tok2006progressive} share similar ideas, but uses a spatial join instead of equi-join. Mihaela et al.,~\cite{bornea2010adaptive} propose a flush policy based on the range of values of  the join attribute and the join result size. Furthermore, Stratis et al., \cite{viglas2003maximizing} consider maximizing output rate for multi-way join and propose a multi-way join operator called MJoin as an alternative to a tree of binary joins. On the other hand, Lawrence~\cite{lawrence2005early} proposes an early hash join algorithm to trade-off between early join results and end-to-end join execution time. The basic idea is through changing reading strategy from the two join tables, e.g., alternative reading or 5:1 ratio from table A and table B. Unlike \cite{tao2005rpj,tok2007rrpj,viglas2003maximizing,lawrence2005early}, PR-join~\cite{chen2010pr} focuses on generating early representative join results with statistical guarantees. PR-join improves the blocked ripple join by adaptively changing the ripple width and achieves a higher early join result rate.}



\section{Conclusions}  

We presented \ntail, a data exploration engine that supports LIMIT queries
by  retrieving \anyk valid records for arbitrary queries as quickly as possible. We
proposed \densitymaps, a lightweight index structure, as well as four \anyk sampling
algorithms built on top of simple cost models. 
Our experimental evaluations demonstrated that \ntail is 
effectively able to trade-off density and locality to 
speed up query runtimes up on average by 13$\times$ on synthetic datasets and 4$\times$ 
and 9$\times$ on real datasets for HDDs and SSDs respectively.

\newpage
\balance
{
\bibliographystyle{abbrv}
\bibliography{main}
}
\techreport{
\appendix
\section{Optimality and Complexity}

\subsection{Optimality Proof for {\large \optimal}}
\label{sec:appendix-proof}
\begin{proof}
We demonstrate here that $Opt(k, \lambda)$ in Algorithm~\ref{alg:optimal} 
gives the optimal I/O cost.
Recall that $C(s,i)$ refers to the minimal cost to retrieve $s$ estimated valid
records when block $i$ is amongst the blocks fetched. 
Our first goal is to verify that $C(s,i)$ satisfies the recursive
equations. 
Let $\Omega$ be the set of selected blocks for $C(s,i)$ 
and $j$ be the block ID just before $i$ in $\Omega$. 
Then $C(s,i)$ will select the same set of blocks from the 
first $j$ blocks as that for $C(s-s_i,j)$, i.e., \mbox{$\Omega \setminus
\{i,j\}$}. 
Otherwise, we can replace one by another to get lower I/O cost. Thus, we have $C(s,i)=\min_{j=1}^{i-1} (C(s-s_i,j)+RandIO(j,i))$ by considering all $j$. Furthermore, since $RandIO(j,i)$ is a constant when $j-i>t$ and $\min_{k=1}^{j}C(s-s_i,k)=Opt(s-s_i,j)$ by considering all the last picked block $k$, we have $\min_{j=1}^{i-t-1} (C(s-s_i,j)+RandIO(j,i))=Opt(s-s_i,i-t-1)+constant$. Thus, the formula can be rewritten as that in the beginning of Section~\ref{sssec:dp}.

First, we observe that $C(s,i)$ has some prefix-optimal property. Let $\Omega$ be the set of selected blocks for $C(s,i)$ and $j$ be the block ID just before $i$ in $\Omega$. Then $C(s,i)$ will select the same set of blocks from the first $j$ blocks as that for $C(s-s_i,j)$, i.e.,$\Omega \setminus \{i,j\}$. Otherwise, we can replace one by another to get lower I/O cost. Thus, we have $C(s,i)=\min_{j=1}^{i-1} (C(s-s_i,j)+RandIO(j,i))$ by considering all $j$. Furthermore, since $RandIO(j,i)$ is a constant when $j-i>t$ and $\min_{k=1}^{j}C(s-s_i,k)=Opt(s-s_i,j)$ by considering all the last picked block $k$, we have $\min_{j=1}^{i-t-1} (C(s-s_i,j)+RandIO(j,i))=Opt(s-s_i,i-t-1)+constant$. Thus, the formula can be rewritten as that in the beginning of Section~\ref{sssec:dp}.
 
Next, we obtain $Opt(s,i)$ by considering two different cases: (a) block $i$ is amongst the blocks fetched; (b) block $i$ is not amongst the blocks fetched. In the first case, $Opt(s,i)$ is exactly $C(s,i)$, while in the second case, $Opt(s,i)$ is exactly the same as $Opt(s,i-1)$. Hence we have the formula stated in the beginning of Section~\ref{sssec:dp}. In all, our proposed DP is correct and $Opt(k, \lambda)$ in Algorithm~\ref{alg:optimal} gives the optimal I/O cost. 
\end{proof}

\subsection{Complexity Analysis}
We analyze the complexity for three of our \anyk algorithms: \thresh, \prong, and
\optimal. Naturally, the complexity of \hybridactual is the maximum of the \thresh and \prong.

\subsubsection{Complexity for {\large \thresh}}
In Algorithm~\ref{alg:threshold},
the set $M$ contains the set of blocks that have already been encountered
by the algorithm, but not yet selected to be part of the output.
We maintain $M$ as a sorted set in the descending order of their densities. 
Therefore, the complexity of insertion (Line 8) is $O(\log(\mid M\mid))$ 
while the complexity of retrieval (Line 10 and 18) and deletion (Line 14) is constant. 
In the worst case, the computational complexity of \thresh is
$O(\lambda \gamma + \lambda \log(\lambda))$, where $\lambda$ is the number of blocks that table
$T$ is allocated on disk. 
However, in practice, the number of predicates of a
query is generally less than $10$, while datasets are usually allocated on
thousands of blocks, indicating we can treat $\gamma$ as roughly constant. Consequently, in general,
the complexity of \thresh is $O(\lambda \log(\lambda))$. Additionally, given that our system
focuses on the cases of browsing $k$ results where $k$ is usually much smaller
than the total number of query records, \thresh usually terminates after
looking into only a small number of blocks instead of $\lambda$ blocks, which
further reduces the computation time in real world scenarios.

\subsubsection{Complexity for {\large \prong}}
The computational complexity of Algorithm~\ref{alg:prong}
consists of two parts; calculating $M$ and performing two (amortized) linear
scans of
$M$: $O(\lambda \gamma + 2\lambda) = O(\lambda\gamma)$.
Typically, $\gamma
< 10$, thus the time complexity can be reduced to $O(\lambda)$.

\subsubsection{Complexity for {\large \optimal}}
The computational complexity of \optimal, shown in Algorithm~\ref{alg:optimal},
is $O(\lambda\gamma + \lambda k t)$. However, we once again apply the fact that
$\gamma < 10$ in practice to reduce the time complexity to $O(\lambda k t)$.

}
\end{document}